\documentclass{aa}
\usepackage{graphicx}
 \usepackage{lscape}
\usepackage{natbib}
\usepackage{txfonts}
\begin{document}
   \title{``TNOs are Cool": A survey of the trans-Neptunian region. XI.}
\subtitle{ A Herschel-PACS\thanks{Herschel is an ESA space observatory
        with science instruments provided by European--led Principal
        Investigator consortia and with important participation from
        NASA. PACS: The Photodetector Array Camera and Spectrometer is
        one of Herschel's instruments.} view of 16 Centaurs}

\titlerunning{Centaurs observed with Herschel-PACS }

   \author{R. Duffard \inst{1}
\and
          N. Pinilla-Alonso \inst{1,2}
\and
           P. Santos-Sanz \inst{1}
\and
         E. Vilenius\inst{3}
\and
          J.L. Ortiz \inst{1}
\and
         T. Mueller \inst{3}
\and
          S. Fornasier \inst{4}
\and
     E. Lellouch \inst{4}
\and
         M. Mommert \inst{5}
\and
         A. Pal \inst{6}
\and
         C. Kiss \inst{6}
\and
         M. Mueller \inst{7}
\and
         J. Stansberry \inst{8}
\and
        A. Delsanti \inst{9}
\and
    N. Peixinho \inst{10}
\and
    D. Trilling\inst{11}
          }

   \institute{Instituto de Astrof\'{i}sica de Andalucia - CSIC, Glorieta de la Astronom\'{i}a s/n, Granada, Spain.\\
              \email{duffard@iaa.es}
         \and
              Department of Earth and Planetary Sciences, University of Tennessee, 1412 Circle Dr, Knoxville TN 37996-1410, USA.
         \and
            Max-Planck-Institut f\"ur extraterrestrische Physik, Postfach 1312, Giessenbachstr., 85741 Garching, Germany.
         \and
             LESIA-Observatoire de Paris, CNRS, UPMC Univ. Paris 6, Univ. Paris-Diderot, 5 place J. Janssen, 92195 Meudon Cedex,
France
         \and
             Deutsches Zentrum fur Luft- und Raumfahrt e.V., Institute of Planetary Research, Rutherfordstr. 2, 12489 Berlin, Germany
         \and
            Konkoly Observatory, Research Centre for Astronomy and Earth Sciences, Konkoly Thege 15-17, H-1121 Budapest, Hungary
         \and
            SRON, Netherlands Institute for Space Research, Low Energy Astrophysics, Groningen, Netherlands.
         \and
           Space Telescope Science Institute, 3700 San Martin Drive,  Baltimore MD 21218. USA.
          \and
           Laboratoire d'Astrophysique de Marseille, CNRS \& Universite de Provence, 38 rue Fr\'ed\'eric Joliot-Curie,
13388 Marseille Cedex 13, France
    \and
    Unidad de Astronom\'{\i}a, Fac. de Ciencias B\'asicas, Universidad de
Antofagasta, Avda. U. de Antofagasta 02800, Antofagasta, Chile.
    \and
    Department of Physics and Astronomy, Northern Arizona University, Flagstaff, AZ 86001, USA.
             }

   \date{Received ; accepted }


  \abstract
{Centaurs are the transitional population between trans-Neptunian objects
(TNOs) and Jupiter-family comets. Their physical properties provide an
insight into TNO properties, but only under restricted conditions since Centaurs
are closer to the Sun and Earth. For this reason it is possible to
access the smaller ones, which is more difficult to do with the TNO population.   }
{The goal of this work is to characterize a set of 16 Centaurs in terms of
their size, albedo, and thermal properties. We study the correlations, for a
more extended sample obtained from the literature, of diameter, albedo,
orbital parameters, and spectral slopes. }
{We performed three-band photometric observations using Herschel-PACS and
used a consistent method for the data reduction and aperture photometry of this
sample to obtain monochromatic flux densities at 70, 100, and 160 $\mu$m.
Additionally, we used Spitzer-MIPS flux densities at 24 and 70 $\mu$m when
available. We also included in our Centaur sample scattered disk objects (SDOs),
 a dynamical family of TNOS, using results previously published by
 our team, and some Centaurs observed only with the Spitzer/MIPS
 instrument.  }
{We have determined new radiometric sizes and albedos of 16 Centaurs. The
first conclusion is that the albedos of Centaur objects are not correlated
with their orbital parameters. Similarly, there is no correlation
between diameter and orbital parameters. Most of the
objects in our sample are dark (pv $<$ 7\%) and most of them are small (D
$<$ 120km). However, there is no correlation between albedo and
diameter, in particular for the group of the small objects as
albedo values are homogeneously distributed between 4 to 16\%. The
correlation with the color of the objects showed that red objects are
all small (mean diameter 65 km), while the gray ones span a wide range of
sizes (mean diameter 120 km). Moreover, the gray objects tend to be darker, with
a mean albedo of 5.6\%, compared with a mean of 8.5\% (ranging from 5 to 15\%)
for the red objects.}
   {}


   \maketitle
%

\section{Introduction}

One fundamental question in astrophysics is how planetary systems form
and evolve.  Accurate physical properties of small solar system bodies
are crucial pieces of information needed to understand the formation
processes, and they constrain models of planetary formation and evolution.
Centaurs are a dynamical class of small bodies in our solar system with
orbits  mostly in the region between Jupiter and Neptune that cross
the orbits of one or more of the giant planets.

The first Centaur was discovered in 1977  and was named Chiron (who is the son of Kronos
and grandson of Uranus in Greek mythology). It was the first minor planet
with a perihelion distance far beyond Jupiter's orbit (Pluto was classified
as a planet at that time). At least two objects were discovered earlier but were only
re-classified as Centaurs after the discovery of Chiron. The next Centaur, Pholus, was discovered only 15 years later. 
Currently, there are 211 Centaurs listed by the Minor Planet Center (MPC) as of early
February 2013
(URL:http://www.minorplanetcenter.net/iau/lists/Centaurs.html). A
considerable fraction of them have been discovered only recently, since the
beginning of 2010. We note that the MPC list includes scattered-disk
objects (SDOs), but that some of those are considered to be Centaurs by some
authors.

Qualitatively, Centaurs are a transitional population between TNOs and
Jupiter-family comets. The exact definition of a Centaur is not generally
agreed upon in the literature.  \cite{Gladman2008} has the most restrictive
definition, which excludes Okyrhoe and Echeclus, which are both considered to
be Centaurs by many. According to the Gladman classification, perihelion distance,
q, and the semimajor axis, a, satisfy a$_J$ $<$ a $<$ a$_N$ and q $>$ 7.35
AU (a$_J$ and a$_N$ are the semi-major axis of Jupiter and Neptune,
respectively). Moreover, the Tisserand parameter needs to be Tj $>$ 3.05 for
these objects. The two mentioned objects are labeled Jupiter-coupled
objects in \cite{Gladman2008}. The most popular definition used in the
literature, which we adopt here, is that $a_{J} <$ a $< a_{N}$
and $a_{J} <$ q $< a_{N}$ \citep{Jewitt2009}. In addition, the object must
not be in a mean-motion resonance with any planet.

Dynamical models suggest that close encounters with planets limit
the median orbital lifetime  of Centaurs to approximately 10 Myr
\citep{Tiscareno2003} and that the lifetime is proportional to the
perihelion distance. Most Centaurs ($\sim$2/3) will be ejected to the
outskirts of the solar system, while the remainder are perturbed into the
inner solar system as short-period comets, broken apart, collide with a
planet, or become a temporary satellite of a planet. Some objects that
transit from a Centaur to a short-period comet and back, or that scatter
back to the Kuiper belt, which is believed to be the main source-region. \cite{Horner2004} found that a Centaur becomes a new Earth-crossing
object every 880 years. Some authors estimated that there are about 44300 Centaurs larger than 1 km in diameter and the estimated
influx from the Kuiper belt is 1/125 yr$^{-1}$ \citep{Horner2004}.
\cite{Horner2006} have shown that Centaurs can be captured as trojans of Jupiter and
other giant planets, that is, they are in a 1:1 mean-motion resonance with the
planet, and that Trojans can also escape and become Centaurs. On the
other hand, \cite{Disisto2007} estimated that SDOs are probably the main
source for Centaurs and provided a far higher estimate, 2.8x10$^8$, for the
current population of Centaurs with R $>$ 1 km.  \cite{Levison1997} suggested
that trans-Neptunian objects in unstable low-eccentricity orbits are a secondary source of Centaurs, with a corresponding population of around 1.2$\times 10^7$.

The surface properties of Centaurs and TNOs are distinct,  probably because of the different influence of surface and dynamical evolution on the two groups of objects. Processes that alter the surfaces include collisions
(especially for bodies $\leq$ 100 km), cometary activity, and space
weather. Many Centaurs seem to have heterogeneous compositions. This can
be explained by fresh areas after impacts or sporadic activity
\citep{Barucci2011}. Due to their rapidly evolving orbits, some
Centaurs may previously have been closer to the Sun and therefore were more
active, even though they are currently inactive. Polarization properties, based
on a small sample of observed Centaurs, indicate that there are distinct
differences in the topmost surface layers of Centaurs compared with TNOs
\citep{Belskaya2010}.

Among the minor planet populations, Centaurs are unique in that their
B-R colors are divided into gray and red populations instead of exhibiting
a continuous range of colors \citep{Peixinho2003}. This bimodal
distribution does not appear in the SDOs, the possible progenitors of
the Centaurs, or in the Jupiter family comets \citep{Jewitt2002}.
\cite{Lamy2009} found no evidence for bimodality either in their Hubble Space Telescope study of the colors
of 51 comets. \cite{Peixinho2003} indicated that only the Centaurs display bimodal colors. On the other
hand, TNOs exhibit a broad continuous color distribution, from neutral/gray to
very red, with no statistical evidence of a color gap between the extrema
\cite[for a review]{Tegler2008}. On the other hand, in a recent work it was
proved that small objects, including both TNOs and Centaurs, display a
bimodal structure of B-R colors at a 0.1\% significance level (i.e. objects
with absolute magnitude H$_R$($\alpha$) $\geq$ 6.8, or D $\leq$165 km
(for an assumed albedo p$_R$ = 9\%), with the ``gap'' centered on B-R =1.60
\citep{Peixinho2012}. \cite{Fraser2012} found that all objects with q$<$35 AU
 (a group that includes all Centaurs) have bimodal colors.

There is no agreed-upon explanation for the observed bimodality of the Centaur colors. Possible explanations include formation in the
presence of a primordial, temperature-induced, composition gradient, or the
influence of comet-like activity (or lack of it). Some Centaurs are
known to exhibit activity, and \cite{Melita2012} explored possible
connections to colors. They found that Centaurs that spend more time
closer to the Sun are more neutral/gray than the others. They suggested
that the neutral/gray colors may result from the formation of a lag deposit of
silicate dust because more volatile ices sublimate during periods of activity.

When only optical photometry data are available, the sizes of distant,
unresolved small bodies can be estimated by assuming a geometric albedo.
When thermal data are also available, the size and albedo can be determined
simultaneously by a radiometric technique (i.e. thermal modeling). There are
only a handful of distant targets, Eris, Makemake, and Quaoar, for instance,
for which a direct diameter estimate via stellar occultation is available
\citep{Sicardy2011, Ortiz2012, Braga2013}.  There is also an occultation diameter for Chariklo (as yet
unpublished).

For objects with occultation diameters and thermal observations, the
two can be combined to provide tight constraints on the temperature distribution
on the surface. The temperature distribution is controlled by factors such
as surface roughness, thermal inertia, and spin vector, so there is the potential
to learn significantly more for these objects than for those with only thermal
or occultation constraints on their diameters.

Several years ago, \cite{Stansberry2008} reported (using the Spitzer
Space Telescope) the results on 15 Centaurs with a geometric albedo in the
range 2.5\% to 18\% and an average of 7$\pm$3\%. The measured diameters were
between 30 to 260 km \citep{Stansberry2008}. Centaurs may have lower albedos
than TNOs on average. Furthermore, based on that sample, Centaurs show a
stronger red color - geometric albedo correlation than TNOs on
average, which has not been explained. More recently, \cite{Bauer2013}
published a set of 52 Centaurs and SDOs observed with WISE. They found a
mean albedo of 8$\pm$4\% for the entire dataset. 

Only very few of the brightest Centaurs have been observed using ground-based mm/submm
telescopes. The Herschel open time key program ``TNOs are Cool!'' observed a
sample of 18 Centaurs. However, that sample significantly overlaps both the
Spitzer sample and the WISE sample (with 17 of the WISE objects overlapping
our Herschel sample). Here we focus primarily on understanding the combined
Herschel plus Spitzer sample of 28 Centaurs, plus an additional 8 SDOs.

In the next section, the observations made with the Herschel Space
Observatory (PACS instrument) are presented, together with a description of
the data reduction. Section 3 describes the thermal modeling applied
to the data. In section 4, the results on the observed sample are presented.
In section 5 we describe an extended sample found in literature.
Sections 6 and 7 present a statistical analysis and the correlations found
in the sample, respectively. We discuss our results in Section 8. Finally, section 9 presents the summary and conclusions.

\section{Observations and data reduction}

Our sample of 18 Centaurs was observed as part of the Herschel key
program ``TNOs are Cool!"  \citep{Mueller2009, Mueller2010}. The data were
collected mainly from March, 2010 to June, 2011. Table 1 presents the
complete list of targets as well as pertinent information on their orbital
parameters, rotational period, and light-curve amplitude, spectral slope, and
whether any ices are known to be present on their surfaces. Data were taken
using the Photodetector Array Camera and Spectrometer (PACS,
\cite{Poglitsch2010}) in the wavelength range 60 - 210 $\mu$m. PACS is an
imaging photometer with a rectangular field of view of 1.75' x 3.5'. The
short-wavelength array has a filter wheel to select between two bands: 60-85
$\mu$m or 85-125 $\mu$m. The long-wavelength band is 125-210 $\mu$m, and
images are collected in that channel simultaneously with either
short-wavelength band. In the Herschel-PACS photometric system these bands
have been assigned the reference wavelengths 70, 100, and 160 $\mu$m and are called blue, green, and red, respectively. For more details
on the acquisition, data reduction, and flux extraction of the data see
\cite{Santos-Sanz2012}. Here we present results for 16 of the 18 Centaurs,
because Chiron and Chariklo were previously presented in \cite{Fornasier2013}.
Finally, the Centaurs 2006 SX$_{368}$ and (42355) Typhon were observed in
the science demonstration phase (two bands only) of the same key project and
were published in \cite{Mueller2010}. 2006 SX$_{368}$ was re-observed in the
routine science phase in all three bands, and the results are presented here. In Table 2 we present 
the identification number of the observation, followed by the duration and mid-time of the
exposure. We also present the heliocentric distance, geocentric distance, and phase angle of the observation.

Of the 18 Centaurs observed with the Herschel Space Observatory, 16 were also
observed by the Spitzer Space Telescope \citep{Werner2004, Gehrz2007}
using the Multiband Imaging Photometer (MIPS, \cite{Rieke2004}). Here
we used Spitzer observations at 24 and/or 70 $\mu$m to complement
the Herschel observations. When the 24 $\mu$m band data are combined with the
Herschel-PACS data, they provide strong constraints on the color temperature
of the target spectrum. When the 70 $\mu$m band data are available, they are mostly
similar to the Herschel data. The Spitzer-MIPS results for most of these
Centaurs were originally published by \cite{Stansberry2008}, but we
also present previously unpublished flux densities for four additional
Centaurs observed under Spitzer program ID 50348 (PI: D. Trilling).
Only two objects from the Herschel-PACS sample were not observed by
Spitzer-MIPS, 2008 FC$_{76}$ and 2002 KY$_{14}$.

All of the MIPS flux densities presented here result from a consistent
reprocessing of all existing Spitzer-MIPS data for TNOs and Centaurs
\citep{Mueller2012}. The reprocessing uses the same reduction techniques
described in \cite{Stansberry2008}, but makes use of updated ephemeris
information for the targets, a key element of the data processing,
particularly for the fainter targets. The flux densities given here (see
Table 3) supersede the values given in \cite{Stansberry2008}. For targets
observed more than once with Spitzer, fluxes are given from each visit, and
the relevant observation identifier (AORKEY) is given in the table notes. 

The Herschel-PACS data reduction from level 0 (raw data) to level 2
(images) was made using the Herschel Interactive Processing Environment
(HIPE\footnote{Data presented in this paper were analyzed using HIPE, a
joint development by the Herschel Science Ground Segment Consortium,
consisting of ESA, the NASA Herschel Science Center, and the HIFI, PACS and
SPIRE consortia members, see
http://herschel.esac.esa.int/DpHipeContributors.shtml}) with modified
scan-map pipeline scripts optimized for the ``TNOs are Cool!" key program.
After identifying the target we measured the flux densities at the
photo-center position using DAOPHOT routines \citep{Stetson1987} for
aperture photometry. A detailed description of how aperture photometry is
implemented in our program is given in \cite{Santos-Sanz2012},\cite{Vilenius2012},\cite{Mommert2012}, and \cite{Fornasier2013}. The absolute photometric accuracy of our
pipeline, based on the photometry of relatively faint standard stars, is
explained in \cite{Kiss2013}. The color-corrected flux densities are given
in Table 3. The uncertainties given there include the photometric
1$\sigma$ and absolute calibration 1$\sigma$ uncertainties.  \\

\section{Thermal modeling}

The main objective of this work is to obtain diameters, albedos, and
surface thermal properties of the targets.  Our targets are too small
to resolve by direct-imaging, but by combining visible, reflected-
light measurements and thermal measurements, it is possible to solve
for the geometric albedo and size, and in some cases also to constrain
the surface temperature distribution. This \emph{radiometric technique}
relies on a model that describes how thermal radiation is emitted from
the surface of the targets, and is  briefly described below.

The flux density of reflected solar light depends on the product of the
target's size and albedo. Using the definition of absolute magnitude $H$,
this is expressed by
\begin{equation}
\label{opt_constraint}
H = m_\mathrm{Sun} + 5 \log \left( \sqrt{\pi} a \right) - \frac{5}{2} \log \left( p S_\mathrm{proj} \right),
\end{equation}
where $m_\mathrm{Sun}$ is the visual magnitude of the Sun, $a$ is the
distance of 1 AU in km, p is the geometric albedo, and $S_\mathrm{proj}$ is
the projected area of the target in km$^2$. $H$, $m_\mathrm{Sun}$ and $p$
are expressed in the same pass-band, usually V or R band. For the majority
of Centaurs, H$_V$ magnitudes were computed using literature
values. For some targets we calculated a new $H_\mathrm{V}$ based on
apparent magnitudes, corrected for the observing geometry, and performed a
linear fit to determine the phase correction.  A few targets do not have
photometric-quality data available; for these we used the Minor Planet Center
data, usually in R-band, together with the average V-R color of Centaurs
from the MBOSS-2 data base.

There are three basic types of models to predict the emission of an airless
and coma-less body with a given size and albedo assuming an energy balance
between insolation and re-emitted thermal radiation: the standard thermal
model (STM) \citep{Lebofsky1986}, the fast-rotator or isothermal-latitude
thermal model (ILM) \citep{Veeder1989}, and the thermo-physical model 
\citep[e.g.]{Spencer1989, Lagerros1996}. The STM assumes a spherical body where
temperatures on the surface depend on the angular distance $\omega$ from the
subsolar point as in a Lambertian emission model:
\begin{equation}
\label{temps}
T \left( \omega \right) = \cos^\frac{1}{4}\left( \omega \right)
\left[\frac{ \left(1-A\right) S_{sun}}{\epsilon_\mathrm{b} \eta \sigma r^2} \right]^{\frac{1}{4}},
\end{equation}
where $A$ is the Bond albedo, $S_{sun}$ the bolometric solar constant,
$\epsilon_\mathrm{b}$ the bolometric emissivity, $\sigma$ the
Stefan-Boltzmann constant, and $r$ the heliocentric distance.  Since the
optical constraint (Eq.~\ref{opt_constraint}) is often used with V-band data
and the temperatures (Eq.~\ref{temps}) depend on $1-A=1-pq$, where $q$ is
the phase integral, we assumed that $A\approx A_\mathrm{V}$ and used q = 0.336
p$_V$+0.479 derived by \cite{Brucker2009} for a sample of TNOs. The
beaming parameter, $\eta$, is an empirical factor that adjusts the
subsolar-point (i.e. maximum) temperature. It approximately accounts for the
combined effects of roughness, thermal inertia, and rotational period. On
rough surfaces heat is radiated preferentially in the sunward direction,
which means a beaming factor $\eta<1$. Values of $\eta > 1$ mimic the
effects of high thermal inertia (or fast rotation).

While the STM discussed above makes the simplifying assumption that the
object has zero thermal inertia, the ILM is the opposite extreme, assuming
infinite thermal inertia. In consequence, the ILM temperature distribution
is isothermal at any given latitude, and represents the coldest end-member
model for the temperature distribution.

As in our previous ``TNOs are Cool'' key program publications
\citep{Santos-Sanz2012, Vilenius2012, Mommert2012}, we used models derived
from the STM: either the Near-Earth Asteroid Thermal Model (NEATM)
\citep{Harris1998} or the hybrid-STM \citep{Stansberry2008}.  Except for the
fact that the hybrid-STM assumes zero phase angle, it is identical to NEATM.
Because all of our targets are observed at phase angles smaller than
$10\deg$, the differences between the two models is expected to be small. However,
thermal phase curves of Centaurs and TNOs are poorly understood, so it is
difficult to quantify exactly how large the differences might be. In the
canonical STM formulation $\eta = 0.76$ \citep{Lebofsky1986}, but in the
NEATM model $\eta$ is adjusted, that is, it is allowed to ``float'', to best fit the
observed emission from each target.

For each target we fit the NEATM to three fluxes (for the Herschel-only
targets), or to four or five fluxes (for the Herschel plus Spitzer targets,
depending on whether only the 24 or the 24 and 70 $\mu$m data were available). If
these floating-$\eta$ fits to the data resulted in un-physical values for
$\eta$ ($< 0.6$ or $> 2.6$), we instead fit the data using $\eta = 1.2\pm
0.35$, based on the average value for TNOs and Centaurs found by
\cite{Stansberry2008}. Based on beaming parameters measured for 85 TNOs and
Centaurs, \cite{Lellouch2013} found median and equal-weight mean values of
$\eta$= 1.09 and 1.175, respectively. Their Figure 1 shows that nearly all
objects at heliocentric distances smaller than 30 AU have beaming parameters in
the range 0.8-1.7, consistent with the range of values we have assumed
above. \cite{Lellouch2013} interpreted the $\eta$ values in terms of a very
low thermal inertia of the TNO/Centaur population.

A simplifying assumption we made (which has been made in
previous publications from our team and from other groups) is that the
nonspherical shape of the objects can be neglected when applying the
radiometric method to derive sizes and albedos. Most TNOs are large enough
to probably be almost spherical (the smaller ones, which might
have irregular shapes, are too faint to have been discovered). Our Centaur
sample, however, includes many objects small enough to be nonspherical, and
several are known to have significant rotational light-curves, as summarized
in Table~1.

To precisely account for the effects of nonspherical shapes on our
results, our modeling would have to account for the exact light-curve phase
of each observation, allowing for the area of the target to be a function of
time. A complication with this approach arises because the measured
rotational periods of some targets are too imprecise to allow for accurate
phasing with our data. Furthermore, our thermal observations typically are 
rather long, so the data from even a single-band observation sample a range of rotational phases. Moreover, our multiband data (even if
only the Herschel data are considered) are acquired at multiple epochs, with
four separate epochs applying at both 70 $\mu$m and 100 $\mu$m, and eight epochs at
160 $\mu$m. The signal-to-noise (S/N) ratio achieved in any single-epoch
observation is typically low because our observations were designed to
achieve a S/N of about 10 when all epochs are co-added. Thus, the
particulars of our Herschel data sets for Centaurs complicate any attempt to
model our rotating targets in a fully self-consistent way as regards their
cross-sectional area as a function of time. Mitigating these complications
somewhat is the fact that modeling of thermal emission from non-spherical
objects shows that thermal light-curves are generally more subdued than they
are in the visible \citep{Santos-Sanz2012}.

To avoid these complications but allow for the effects of
nonspherical shapes on our results, we systematically applied larger
uncertainties to our H magnitudes, as described in \cite{Vilenius2012}. For
the objects with measured light-curves, we added 88\% of the half-amplitude
quadratically to the formal uncertainty on the measured H magnitude. For
objects with unknown light-curve amplitude, we assumed a light-curve
half-amplitude of 0.2 mag. For objects such as Pholus and Asbolus, with
strong light-curves, this results in an effective uncertainty of 0.3 -- 0.4
mag. By making this adjustment to the H magnitudes we used in our modeling,
the error bars we derived on the albedo, diameter, and beaming parameter
are expected to encompass the actual values for our targets.

\begin{figure*}
 \centering
   \includegraphics[width=20cm]{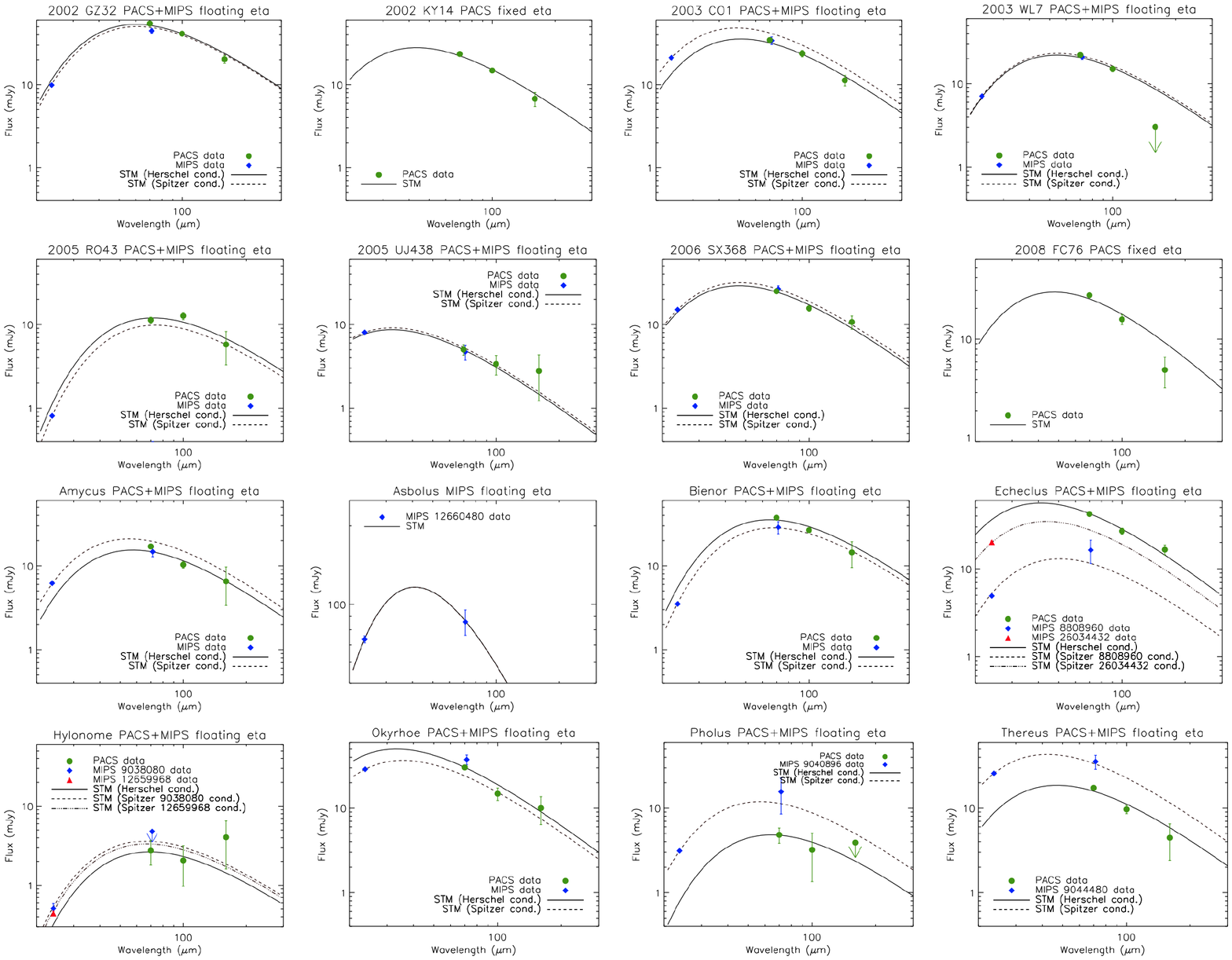}
\caption{Thermal modeling of all the Herschel targets. For PACS+MIPS plots
the filled line is the best-fit (hybrid-STM) for the geometrical Herschel targets, at the heliocentric distance
r(h) and geocentric distance $\Delta$ conditions. The dashed line is the best-fit using the same
model but for the Spitzer geometrical conditions. As discussed in the text,
the preferred solution for Pholus is the one with AORKEY 12661760. }
         \label{Fig1}
   \end{figure*}

\section{Results using Herschel-PACS}

All the measured fluxes using Herschel-PACS and the new (updated or
previously unpublished) fluxes from Spitzer are included in Table 3
together with the absolute magnitude, diameter, albedo and $\eta$ from the thermal modeling. The error estimates of the geometric albedo, the
diameter and the beaming parameter were determined by a Monte Carlo method, as
described in more detail on \cite{Mueller2010}, \cite{Santos-Sanz2012}, and \cite{Mommert2012}.

In this section we describe each target and also include information
on the spectral signatures that give clues on the ices present on the
surface of the targets. The  information on the overall spectral shape in
the visible and near-infrared is presented using the Barucci taxonomy
\citep{Barucci2005} and updates on that work presented in
\cite{Fulchignoni2008} and \cite{Perna2010}.

Objects with a neutral/gray color with respect to the Sun are classified as the BB
(blue or gray) group, and those with a high red color are classified as RR
(red). The BR group consists of objects with an intermediate blue-red color,
while the IR group includes moderately red objects. In our analysis we used the spectral slope in the visible that has been widely interpreted as an
indication of surface composition. Steep red slopes have typically been 
associated with complex organics (tholins) on the surface \citep{Cruikshank2005}. Neutral/gray slopes, on the other hand, are related to
highly processed surfaces covered by dark carbonaceous materials
\citep{Andronico1987}  or high-albedo water-ice-rich surfaces
\citep{Pinilla2008}. All these variables are presented in the last two
columns of Table 1.

Centaurs are mainly distributed in the BR and RR classes, with a
similar H$_2$O-ice-content distribution. There are no Centaurs with an abundant surface-ice content, that is, higher than 20\%
\citep{Barucci2011}. The majority of Centaurs observed multiple times
have an heterogeneous composition. This seems to be the main characteristic of the Centaur population: the variation that affects the new areas that surfaced or were altered by impacts while the Centaurs were still in the transneptunian region. This variation may also have been caused by temporal or sporadic activty.

\subsection{Results on individual targets}

In this subsection we present the results and all the information we
consider important for modeling and interpreting the results
for each individual object. The rotational period, which is useful in thermo-physical
models, and the light-curve amplitude, which one needs to know to determine how spherical the object is
or how inhomogeneous the surface, are also presented in Table 1.
Information on the spectral slope is also presented and included in the
same table. Here we give information on the spectral signatures, which give
clues on the ices present on the surface of the targets. The spectral
information (visible spectral slopes) is used below to determine the
correlation between size/albedo and surface composition. If cometary
activity was reported on the object it is indicated here.

{\bf (95626) 2002 GZ$_{32}$}: has a BR spectral type with a
tentative indication of water ice in the surface
\citep{Barucci2008}. Our results indicate that it is the largest
object of the sample with D= 237$\pm$8 km, a low albedo of 3.7$\pm$0.4
\%, and $\eta$=0.97$^{+0.05}_{-0.07}$. We used Spitzer-MIPS data to
fit a thermal model with five points. No cometary activity was reported
in the literature. The light-curve indicates a rotational period of
5.80$\pm$ 0.03 hours with a low amplitude of 0.08$\pm$0.02 magnitudes
\citep{Dotto2008}. This object is spectrally gray with a moderately small
spectral gradient of 8\%/$10^3$ \AA. This object has an indication of a
weak absorption band centered around 4300 $\AA$ (wide 200 $\AA$ and a depth
of about 3\% with respect to the continuum). If real, this feature would
be similar to the absorption at 4300 $\AA$ found on some primitive main-belt asteroids and attributed to a ferric iron Fe$^{3+}$ spin forbidden
absorption in minerals derived from the aqueous alteration
process such as iron sulfate jarosite \citep{Fornasier2004}.  \\

{\bf (250112) 2002 KY$_{14}$}: also known as 2007 UL$_{126}$, has RR spectral type with an indication of water ice in the
surface \citep{Barucci2011}. \cite{Thirouin2010} reported a rotational
one-peak period of 3.56/4.2$\pm$0.05 hours with an amplitude of
0.13$\pm$0.01 magnitudes. In our results we tried to model it with
a floating $\eta$ but the fit was not acceptable, so we applied a
fixed $\eta$ model.  From the fixed $\eta$ of 1.20$\pm$0.35 our
preferred solution is a diameter of 47$^{+3}_{-4}$ km and an albedo
of 5.7$^{+1.1}_{-0.7}$\%. Only the Herschel-PACS fluxes were fitted because
there are no Spitzer observations of this object.  \\

{\bf (281371) 2008 FC$_{76}$}:  \cite{Fornasier2009} reported a
spectral slope of 36.0$\pm$0.7\%/10$^3$ $\AA$. The spectral type is RR with no detection of water ice or any other ices in the surface
\citep{Barucci2011}.  Its rotational period is not known yet. No activity
was observed in this object. Our results indicate that the object has a
68$^{+6}_{-7}$  km diameter with an albedo of 6.7$^{+1.7}_{-1.1}$ \% .
This object was modeled with only the three Herschel-PACS fluxes, using a
fixed $\eta$ model.  \\

{\bf (136204) 2003WL$_7$}: has a BB spectral type
\citep{Barucci2011}. Light-curve information is provided by
\cite{Thirouin2010}, who reported a single-peak rotational period of 8.24
$\pm$0.05 hours with an amplitude of 0.05 $\pm$0.01 magnitude. The
fitted model to the five points (Herschel + Spitzer) gives a diameter
of 105$^{+6}_{-7}$ km, an albedo of 5.3$\pm$1\%, and an $\eta$ =
1.02$^{+0.07}_{-0.05}$. The flux at 160
$\mu$m is an upper limit and is below the fitted model.  \\

{\bf 2005 RO$_{43}$}: there are no works in the literature that describe the
physical properties of this object. In our case, the fitted model to the five
points (Herschel + Spitzer) gives a diameter of 194$\pm10$ km, an albedo
of 5.6$^{+3.6}_{-2.1}$\%, and an $\eta$=1.12$^{+0.05}_{-0.08}$. There is
an excess in flux at 100 $\mu$m (see Fig. 1). \\

{\bf (145486) 2005 UJ$_{438}$}: this is the smallest object of
our sample. Our results using a free $\eta$=$0.34^{+0.09}_{-0.08}$
indicate that the object has a 16$^{+1}_{-2}$ km diameter with an albedo
of 25.6$^{+9.7}_{-7.6}$ \% . This surprising low $\eta$ could be indicative of activity. In some sense
the high albedo is a direct effect of the low $\eta$. The object was
at the detection limit, around 5-6 mJy at PACS wavelengths. 
The results on this object presented by \cite{Bauer2013} also included
a high albedo, in this case obtained through a fixed $\eta$ model.
\cite{Lellouch2013} also mentioned that this object has
an extremely low $\eta$ value. This might result from coma activity, because emission of small and hot particles would enhance the short-wavelength flux, leading to a low apparent $\eta$.\\

{\bf (248835) 2006 SX$_{368}$}: in a previous work of the TNOs are Cool!
program \citep{Mueller2010} we reported a fixed $\eta$-approach modeling
that yielded a diameter of 79$\pm$9 km and p$_v$ = 5$\pm$1 \%. This object is BR
in the taxonomy. It is on a very eccentric orbit, near the 5:4 mean motion resonance with
Uranus. In \cite{Mueller2010} the Herschel-PACS 70 $\mu$m detection was
compatible with diameters in the range of 70-80 km and an albedo of 5-6\%,
consistent with our results. The upper 160 $\mu$m  flux limit  in that
work constrains the diameter range to values below 105 km and an albedo
higher than 3\%, both in agreement with the measured 70 $\mu$m-flux. No
light-curve is published for this objects. \cite{Jewitt2009} reported this
object as active. \cite{Fulchignoni2008} classified it as BR type. Recently, \cite{Perna2013} searched for signs of activity in this object,
with negative results, which sets limits on the dust production rate.  
Our results, including the Spitzer values, gives a diameter of 76$\pm$2 km,
with an albedo of 5.2$^{+0.7}_{-0.6}$\%, and a final floating $\eta$ =
0.87$^{+0.04}_{-0.06}$ .\\

{\bf (120061) 2003 CO$_1$}: the spectral type of this object is BR
\citep{Perna2010}.  Analysis of the light-curve of this object in two
different works resulted in two different estimate for the rotational
period: 10.00$\pm$0.01 hours with an amplitude of 0.10$\pm$0.05
\citep{Ortiz2006}, and a single-peak rotational period of 4.51$\pm$0.05
hours with an amplitude of 0.07$\pm$0.01 magnitudes reported by
\cite{Thirouin2010}. There is a tentative detection of water ice in the
surface \citep{Barucci2011}. This object has Spitzer observations that
combined with those of Herschel-PACS give five points to be fitted with
the thermal model. The final results give a diameter of 94$\pm$5 km,
an albedo of 4.9$^{+0.5}_{-0.6}$\%, and $\eta$=1.23$^{+0.12}_{-0.11}$.\\

{\bf Amycus}: the reported spectral type is RR with a tentative detection
of water ice in the surface \citep{Barucci2011}. Our results indicate
that the object has a 104$^{+8}_{-8}$ km diameter with an albedo of
8.3$^{+1.6}_{-1.5}$\% and $\eta$= $1.00^{+0.12}_{-0.13}$.  The thermal
model fit presents a flux excess at 24 $\mu$m, which may be related
with two different albedo terrains, as was suggested in
\cite{Lim2010} for Makemake.  The light-curve gives a rotational period of
9.76 hours with an amplitude of 0.16 magnitudes \citep{Thirouin2010}. The
compositional spectral model indicates amorphous carbon, Triton tholin
and water ice in the surface \citep{Doressoundiram2005}, which is compatible with
the RR type. \\

{\bf Bienor}: is a BR spectral type Centaur with a positive detection
of water ice in the surface \citep{Barucci2011}. Our results indicate
that this object has a 198$^{+6}_{-7}$ km diameter with an albedo of
4.3$^{+1.6}_{-1.2}$\% and a $\eta$ = 1.58$^{+0.07}_{-0.07}$. The reported
rotational period is 9.14 hours  with an amplitude of 0.75 magnitudes,
or 9.174$\pm$0.001 hours with an amplitude of 0.34$\pm$0.08 magnitudes
\citep{Ortiz2002, Ortiz2003}. \\

{\bf Echeclus}: There are two different Spitzer-MIPS observations, but
the fitted model gives compatible results. The preferred result is the
one using  the two Spitzer observations combined with the Herschel observation. Our results indicate that the object has a 64.6$\pm$1.6 km diameter with an albedo of 5.20$^{+0.70}_{-0.71}$\%
using an $\eta$=0.861$^{+0.037}_{-0.036}$. Figure 1 shows the fit with MIPS AORKEY 26034432 and 8808960  and the Herschel observation. No light-curve information is present in the literature.

Neither is there a spectral signature indicating ices on the
surface \citep{Guilbert2009}.  This object presented a peculiar cometary
activity. The source of cometary activity appears to be distinct from Echeclus
itself. The brightness distribution of this source does not follow that
of a cometary coma created by a point-like source (cometary nucleus). It
was reported to look like a diffuse source \citep{Rousselot2008}.  \cite{Bauer2008}
suggested -based on Spitzer imaging- that this Centaur had sustained
activity. Echeclus has
recently turned on again (IAU Circular 9213), which makes it very likely
that its cometary activity is related to volatile sublimation or water-ice
crystallization and not triggered by an unknown external phenomenon. \\

{\bf Hylonome}:  in this case, there are also two different observations
with Spitzer.  The adopted value was the combination of these two
observations with that of Herschel one. Our results indicate that the object has a 74$\pm$16
km diameter with an albedo of 5.1$^{+3.0}_{-1.7}$\% and an $\eta$
= 1.29$^{+0.31}_{-0.31}$ . No light-curve information is present in
the literature. Figure 1 only shows the fit with one of the MIPS
AORKEY 9038080. The fit with the other MIPS AORKEY is very similar.   \\

{\bf Okyrhoe}: our results indicate that the object has a
35$\pm$3 km diameter with an albedo of 5.6$^{+1.2}_{-1.0}$\% and
$\eta$=0.71$^{+0.12}_{-0.13}$ . No light-curve information is available in
the literature. \cite{Barucci2011} reported a tentative detection of water
ice in the surface and classified this object as a BR in the taxonomy. \\

{\bf Pholus}: our preferred results is the one using Spitzer AORKEY 9040896 and Herschel observations because the other Spitzer AORKEY has no 70 $\mu$m detection. The object has a
99$^{+15}_{-14}$ km diameter with an albedo of 15.5$^{+7.6}_{-4.9}$\% and
an $\eta$= $0.77^{+0.16}_{-0.16}$. This object has a reported rotational
period of 9.98$\pm$0.01 hours with different amplitudes ranging from
0.15 to 0.60 magnitudes \citep{Buie1992, Hoffmann1992, Perna2010}. The
reported spectral class is RR \citep{Perna2010}. The spectral information
shows a clear indication of water ice in the surface \citep{Barucci2011,
Guilbert2012}. There is also an indication of methanol (CH$_3$OH) in
the spectra of this object \citep{Barucci2011}.  \\

{\bf Thereus}: our results indicate that the object has a 62$^{+3}_{-3}$ km
diameter with an albedo of 8.3$^{+1.6}_{-1.3}$\% and an
$\eta$=$0.87^{+0.08}_{-0.08}$ . The reported rotational period is 8.33$\pm$
0.01 hours from different sources \citep{Ortiz2002, Ortiz2003,
Rabinowitz2007, Brucker2008}. The light-curve amplitude is 0.16-0.18 or its
double 0.34-0.38.  Its classification is in the BR taxonomy type.
\cite{Licandro2005}  detected spectral rotational variation and modeled
the surface with amorphous carbon silicates, tholins, and
different water-ice quantities.   \\

{\bf Asbolus}: the preferred solution considered only
MIPS 12660480 observations. We discarded MIPS 9039360 because the Spitzer
fluxes are probably affected by activity.  There was a great change in heliocentric distance between both MIPS observations. We
tried to fit the two MIPS fluxes separately with the Herschel-PACS fluxes
without acceptable result. The selected solution was the one with the
strongest signal, that is the one with the shorter heliocentric distance. The
adopted solution gives an estimated diameter of $85^{+8}_{-9}$ km, an
albedo of $5.6^{+1.9}_{-1.5}$\%, and an $\eta$ = $0.97^{+0.14}_{-0.18}$. \\

\begin{figure}
   \centering
   \includegraphics[width=\columnwidth]{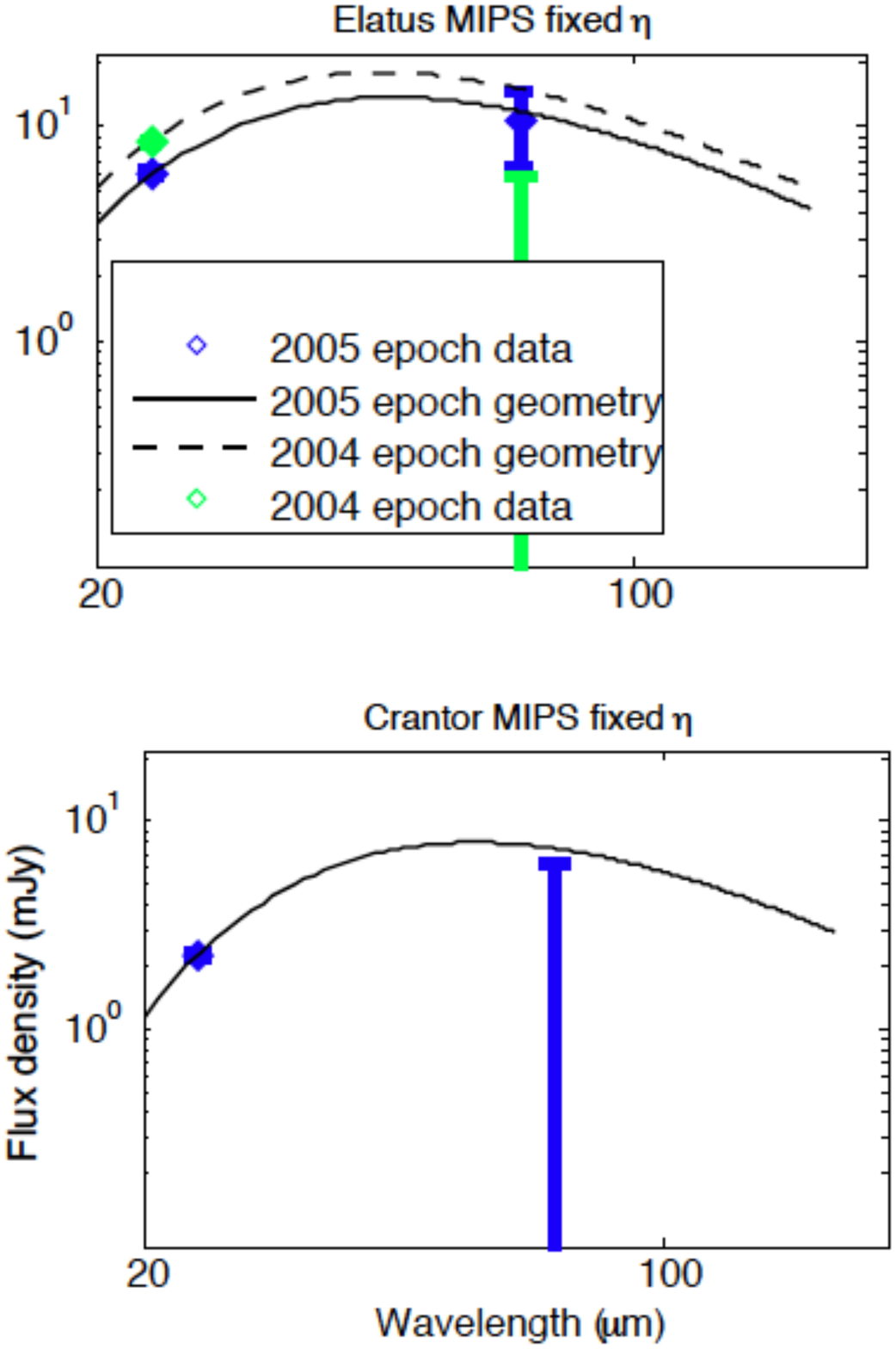}
      \caption{Thermal modeling of two of the new or re-modeled
 Spitzer-only observed Centaurs. Only these two targets had two-band
 data. In the particular case of Crantor the upper limit at 60 $\mu$m was
 treated as a data point with zero flux and 1$\sigma$ error-bar, equaling
 the upper limit value (0$\pm$6.31 mJy). The model for Crantor also considers a fixed $\eta$. The other Spitzer-only objects
 are modeled with only one point at 24 $\mu$m and with fixed $\eta$.
              }
         \label{Fig2}
   \end{figure}

Finally, there is a list of Centaurs observed with Spitzer-MIPS, some of
them published in \cite{Stansberry2008}. Here, we used updated values for
Nessus, Elatus, Cyllarus, Crantor, and 2001 BL$_{41}$. Moreover, we included
in the analysis four objects observed with Spitzer-MIPS that have not yet
been published: (119315) 2001 SQ$_{73}$, 2002 VR$_{130}$, 2004 QQ$_{26}$,
and 2000 GM$_{137}$.  Model results for the two objects observed only
with Spitzer-MIPS (updated and unpublished) are shown in Fig \ref{Fig2}.

The final sample of Centaurs we analyzed (Fig. 1
and Fig. 2) consists of 14 objects observed with Herschel-PACS and
Spitzer-MIPS, 2 Centaurs observed only with Herschel-PACS (2008 FC$_{76}$
and 2002 KY$_{14}$) and 9 objects observed only with Spitzer-MIPS
(including four that were previously unpublished).

\section{Extended Centaur sample observed with Herschel}

Here we describe the radiometric diameters and albedos for Centaurs derived
from the literature. We combine these results with those presented above for
the statistical and correlation analysis presented in the next section.

Chiron and Chariklo were observed with the Herschel PACS and SPIRE
instruments at 70, 100, 160, 250, 350, and 500 $\mu$m. The results were
presented in \cite{Fornasier2013} and are summarized below along with other
results.

{\bf Chiron:} this is the first discovered Centaur and the first with
detected activity. Some changes in the intrinsic brightness were detected by
Hewitt and Bowell in 1978, described in \cite{Bus1989}, and the sudden
brightening of Chiron between 1988 and 1989 \citep{Tholen1988, Bus1988,
Hartmann1990} confirmed the cometary nature of this object. The coma around Chiron was first detected by \cite{Meech1989} when it was at a
heliocentric distance of 11.8 AU. Both long-term (months to years) and
short-term (hours to days) variations in Chiron's absolute magnitude have
been identified and are attributed to ongoing, episodic cometary activity.
These variations were analyzed in detail by \cite{Duffard2002} and updated
more recently by \cite{Belskaya2010}. The variation of H$_V$ over nearly 40
years shows several peaks of activity, with the absolute magnitude changing
by as much as two magnitudes as a result of cometary activity.

In addition to the spectral signatures attributed to water ice on
its surface \citep{Foster1999}, Chiron is one of the largest known Centaurs.
However, estimates of its size vary considerably (perhaps because the
contribution of the dusty coma has not been accounted for). Using various
techniques the diameter of Chiron has been estimated to be 180 km
\citep{Lebofsky1984}, 372 km \citep{Sykes1991}, and 182-189 km
\citep{Campins1994, Marcialis1994}.  \cite{Fernandez2002} found a
diameter of 148$\pm$8 km with an albedo of 0.17$\pm$0.02 and
\cite{Groussin2004} found a radius of 142$\pm$10 km.  

More recently, \cite{Fornasier2013} analyzed Herschel PACS and SPIRE data
for Chiron (assuming H$_V$ = 5.92 $\pm$ 0.20). Based on their thermal
modeling, Chiron's diameter is 218$\pm$20 km, and its geometric albedo is
16 $\pm$3\%.  These authors considered the effect of Chiron's
coma in some detail  and estimated that at the time of the Herschel observations the coma
contributed no more than 4\% to the observed fluxes. There was no evidence
of a resolved coma in these data, and the authors were able to set an upper limit
on the dust production rate (based primarily on the PACS 70 $\mu$m data).

{\bf Chariklo:}  The Spitzer telescope observed Chariklo at 24 and 70 $\mu$m
using MIPS. Chariklo was also observed with the Widefield Infrared
Survey Explorer (WISE)  at 11.6 and 22.1 $\mu$m \citep{Wright2010}. \cite{Fornasier2013} combined Herschel observations
with those of Spitzer and WISE.

Assuming an H$_V$ = 7.40$\pm$0.25 magnitude corresponding to the
absolute magnitude estimate that is closest to the observations (March-June
2008, from Belskaya et al. 2010), the diameter derived from the NEATM model
of the revised Spitzer-MIPS and Herschel PACS and SPIRE data is D =
236.8$\pm$6.8 km and a geometric albedo of 3.7$\pm$1\%. Including WISE
data, the preferred solution using the TPM method and all the observed
wavelengths is D = 248$\pm$18 km and an albedo of 3.5$\pm$1\%.

The fact that the TPM model excludes the pole-on solution at the
time of the Herschel observations reinforces the assumption made by
\cite{Belskaya2010}, who speculated that the 2007-2008 observations
after the Chariklo passage at perihelion (corresponding to a fainter
Hv magnitude) were made viewing the equator, while the 1999-2000
observations, which show a higher Hv value and short-term brightness
variations, corresponded to a pole-on geometry. For the surface
composition, \cite{Dotto2003} reported the spectral signature
for water ice in the surface of Chariklo.

An important conclusion of \cite{Fornasier2013} is that both
Chiron and Chariklo show a significant decrease of emissivity for
wavelengths $\geq$ 100 $\mu$m. The decrease in emissivity at very long
wavelengths probably means  that at these wavelengths  the top surface layer
becomes more transparent and we see the subsurface (which has lower temperatures and higher thermal inertias).   \\

There are two other Centaurs observed previously with Herschel. As reported
in \cite{Mueller2010}, Typhon was observed in the science demonstration
phase; additional observations were presented in \cite{Santos-Sanz2012}, who
included it in their sample of SDO's. The object 2006
SX$_{368}$ was observed in the science demonstration phase. These data have
not  been previously published, therefore we analyzed them and include the
results here.

{\bf Typhon}: \cite{Mueller2010} reported on Herschel science demonstration
phase observations of the binary Centaur, Typhon, finding D = 138$\pm$9
km, p$_V$ = 8$\pm$1\%, and $\eta = 0.96\pm0.08$. Their analysis was based
on both Spitzer and Herschel fluxes. The object has a rotational period
of 9.67 hours and a light-curve amplitude of 0.07$\pm$0.01 magnitudes.

Typhon was also observed in the science routine phase in the three
Herschel-PACS bands. Results were published in \cite{Santos-Sanz2012},
presenting a diameter of 185$\pm$7 km, an albedo of 4.4$\pm$0.3\%,  and an
$\eta$ value of 1.48$\pm$0.07. As mentioned in \cite{Santos-Sanz2012},
the SDP observations were performed in chop-nod mode (less sensitive
than scan-map for point-sources), which gives lower 70 and 100 $\mu$m
fluxes and no detection at 160 $\mu$m. In this work we prefer the results of 
\cite{Santos-Sanz2012}.   Typhon is a Centaur according to the MPC and
an SDO according to the Gladman definition. We include this object in the
final complete (Centaurs + SDOs) sample.

By combining their Herschel diameter for Typhon with the system mass
determined by \cite{Grundy2008}, \cite{Santos-Sanz2012} derived a bulk
density of 0.66$^{+0.09}_{-0.08}$ g cm$^{−3}$. This is slightly higher
than the value reported by \cite{Grundy2008}, who used the diameter reported by
\cite{Stansberry2008}, based on a single-band Spitzer detection.  \\

\section{Statistical analysis}

In this and the next section we analyze our results on Centaurs and search for
correlations between the albedo and the diameter and some of their physical
characteristics, such as orbital parameters and visible spectral slope. When
we do not specify otherwise, we refer to the whole sample of Centaurs together
with the sample of SDOs.

\begin{figure}
   \centering
   \includegraphics[width=\columnwidth]{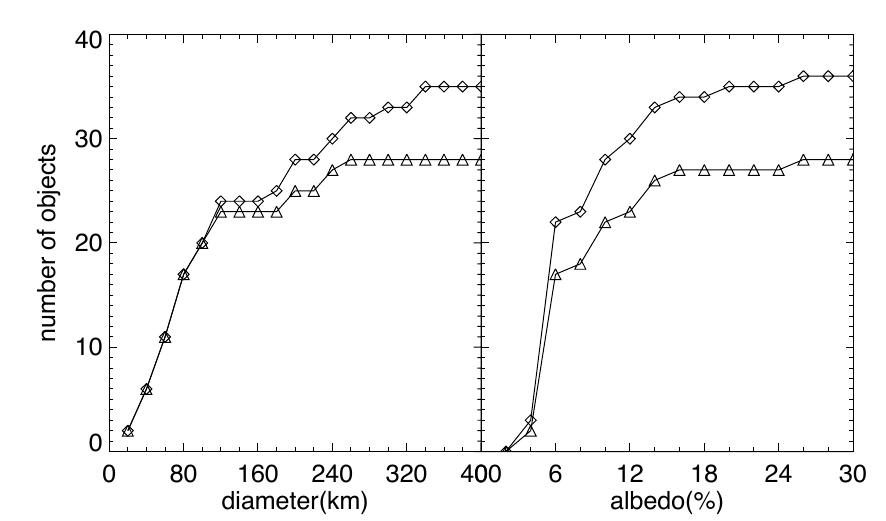}
\caption{Cumulative distribution for diameters and albedo for the whole
sample (Centaurs + SDOs, squares) and Centaurs only (triangles). In the
left panel 2007 OR$_{10}$ is excluded because its diameter (D=1280 km) is beyond the scale.
              }
         \label{Fig3}
   \end{figure}

The geometric albedos for our sample are mostly lower than 13\% with two
exceptions, 2005 UJ$_{438}$ with 25.6$^{+9.7}_{-7.6}$\% and Chiron
with an albedo of 16$\pm$3\%  (the only one considered active at the time of
observation by Herschel). The albedo of 25\% for 2005 UJ$_{438}$ and the
$\eta$ = 0.34$\pm$0.09 indicate that this object might be active or has
been active recently. The data indicate a color temperature of
$\sim$161 K. This is higher than the instantaneous subsolar temperature of a
zero-albedo object at the relevant heliocentric distance, 136 K (hence the
low equivalent $\eta$ value that is obtained within a NEATM). Therefore this
high color temperature might be due to superheated small dust grains. 
This requires the grains to be small; otherwise they would have the same
temperature as the nucleus.  The darkest object is Chariklo with an albedo
of 3.5$\pm$1\% followed by 2002 GZ$_{32}$ with an albedo of 3.7$\pm$0.4\%.
The mean albedo for our sample is 6.9$\pm$4.8\% for the Centaur population
and 6.7$\pm$4.8\% considering the entire Centaur and SDO sample. As can be
seen in Fig. \ref{Fig3} (right plot), there is a concentration of objects
with albedos between 2 and 6\% that comprises 61\% of the sample. The other
39\% show more disparate values between 6 and 26\%. Including SDOs in the
sample yields similar results, with 65\% concentrated in the 2 - 6\% range
(see Fig. 3).

In the Centaur sample (Herschel sample) the largest object is 2002
GZ$_{32}$ with a diameter of 237$\pm$8 km, and the smallest is 2005
UJ$_{438}$ with a diameter of 16$^{+1}_{-2}$ km. When we include the
Spitzer sample, the smallest Centaur is 2000 GM$_{137}$ with 8$\pm$1.5 km
in diameter. In the whole sample, 82\% of the objects are smaller than
120 km in diameter. SDOs in our sample are consistently larger than the
Centaurs, only one is smaller than 120 km (2002 PN$_{34}$ with a diameter
of 112$\pm$7 km). This is a clear discovery/observational bias. In Fig. \ref{Fig3}, left panel, we show that the distribution of sizes in the sample of Centaurs and Centaurs plus SDOs is not continuous, there is a gap from 120 to 190 km where we can find only one object,  2002 XU$_{93}$ an SDO with a diameter of 164$\pm$9 km.  71\% of the objects are smaller than 120 km. 

The separation into small and large objects in the Centaur and TNO
population is not new. The small objects in these populations, those
with a diameter $<$150 km, are thought to be fragments from collisions
of larger ones. This is an interesting point because their surfaces may have been remodeled by the impact. \cite{Peixinho2012} stated a limit of
165 km using a mean albedo of 9\%. In our sample we have no objects
between 120 km and 190 km, therefore we took 120 km as the limit to distinguish
between small and large objects. The mean albedo for our small and
large Centaurs is 7.0$\pm$4.0\% and 6.2$\pm$5.0\%, respectively. They
are not distinguishable within the errors which suggests that there is
no clear dependence of the albedo and size of this objects. We
study this and other correlations in the next section.

Following what is shown in Fig. 3 that most of the centaurs are small, and most of the centaurs are dark, one can ask here whether these two “most common” populations in the size and albedo distribution (the small Centaurs and the low-albedo objects) are related. Are the most abundant small Centaurs also the darkest? Figure 4 clearly shows that this is not true, because the albedo of the small Centaurs takes any value from 4\% to 14\%. That interval represents almost the entire range of albedos for the whole population (4\% - 16\%). Qualitatively, the albedos of the small objects are not different from the albedo of the whole sample (including the scattered objects), with the exception of 2005 UJ$_{438}$ which we commented above.

Recently, Bauer et al (2013) published a statistical analysis of
52 centaurs and SDOs observed with WISE. Our results are similar
to theirs. There are 17 objects in common in both samples. Both the
diameters and the albedo estimate are similar within the error
bars, discarding any systematic deviation of one set of results with respect
to the other. For most of the sample, our results are more precise therefore,
even if our sample is smaller than theirs, our results complement the
previous ones because they provide better estimate of both diameter
and albedo.  However, there are some particular objects where the results
are clearly different above the 1$\sigma$ limit. The diameter of Thereus is
larger for the WISE fit (86.5$\pm$1.9 km) than our result (62$\pm$3km). The
albedos estimated for Chariklo, 2002 KY$_{14}$, and 2008 FC$_{76}$ are 7.5$\pm$1.5\%, 
18.5$\pm$4.6\% and 12.0$\pm$2.7\% using WISE data, these values are higher
than our estimate (3.5$\pm$1.0\%, 5.7$\pm$1.0\%, and 6.7$\pm$1.5\%, respectively).

Different factors might be the origin of these difference such us the use of a different value of the absolute magnitude, or fits
of the fluxes using only one or two bands instead of four for the WISE data, or two fluxes instead of three for the Herschel
data. However, we investigated these effects and there is no consistent explanation for these differences.
We believe that a combination of these
and other factors, such as the S/N of the observations, might be the cause of
the disparity. The case of Amycus is different. Both results are
compatible within the errors, either the diameter or the albedo, but
the error bar is much larger for WISE (by a factor of 5 for
the diameter and $\sim$10 for the albedo), which may indicate that there is a problem with the WISE data or the reduction for this object. Our result (D=104$\pm$8 km and pv=8.3$\pm$1.5\%) is probably more reliable for
this object.  These differences affect the average values slightly. Bauer found a mean albedo of 8$\pm$4\%, we found a similar value of 6.9$\pm$4.8\% for the Centaurs and 6.7$\pm$4.8\% for the whole sample.

\section{Correlations}

Correlations of albedo, diameter, and beaming parameter with parameters
such as inclination i, eccentricity e, semi-major axis a, perihelion
distance q, aphelion distance Q, and heliocentric distance at the moment
of observation, r(h), (e.g. orbital parameters, visible spectral slope)
have been discussed in previous studies of TNOs.  We searched for such
correlations within our sample using a modified Spearman rank correlation
test, taking into account symmetrical and asymmetrical error-bars
\citep{Spearman1904, Peixinho2004, Santos-Sanz2012}.  Roughly speaking,
a correlation is said to be strong when $| \rho | >0.6$. However, the
evidence for a correlation is given by the significance
level, or equivalently by the confidence level, not by the value of
the correlation coefficient. That is, if we measure a correlation of
$\rho=0.8$, but achieve a significance of only $0.2$ (i.e. a $20\%$
probability of being a random effect), the correlation is considered
to be statistically insignificant.  Table 4  shows correlation coefficients
and the corresponding significance values and confidence levels for all of
the parameter combinations we have examined. The strongest correlations
are highlighted  in boldface.

To increase the sample, we also considered diameters and albedos
of the eight SDOs from \cite{Santos-Sanz2012}. We analyzed the results as the
whole sample (Centaurs+SDOs) and separately (Centaurs only). We also
defined a third group corresponding to objects smaller than 120 km,
based on the findings of \cite{Peixinho2012} and \cite{Fraser2012}. Because 2007 OR$_{10}$
lacks a published visible spectrum, but has an NIR spectrum similar to that
of Quaoar \citep{Brown2011}, we assumed that their visible spectral slopes
are the same.

The only strong correlation found for the Centaurs is between diameter
and heliocentric distance (figure 5, bottom). This correlation persists
and is even stronger when we added the SDOs to the sample. When we considered only the
smaller Centaurs, this correlation is weaker but still present. This is
probably an observational bias that we discuss below.

When we consider the full sample, the correlation between diameter and
semi-major axis, perihelion distance, and inclination are strong.
However, the correlation between diameter and eccentricity is weak for
the full sample, and that between diameter and eccentricity becomes weaker for the Centaur sample alone.

For the smaller Centaurs there are only weak correlations, such as for
albedo versus spectral slope, diameter versus semi-major axis, and diameter
versus heliocentric, perihelion, and aphelion distances. We note that for
the Centaurs, whit significant orbital eccentricities, all of these
distances are correlated, so their correlations with the physical parameters
should be, and are, similar.

\begin{figure}
   \centering
   \includegraphics[width=\columnwidth]{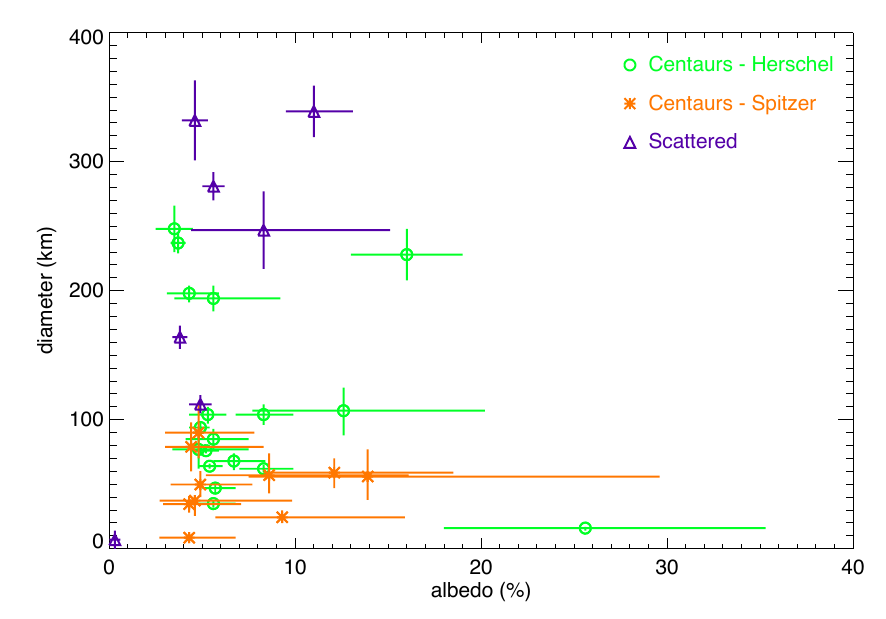}
\caption{Diameter versus albedo plot. We can see the gap in size between
120 to 190 km for the Centaurs. Objects with an albedo higher than 15\% are
suspected to be active. 2007 OR$_{10}$ is not included because its diameter
(D=1280 km) is beyond the scale.
              }
         \label{Fig4}
   \end{figure}

We do not find evidence for a correlation between albedo and diameter
in our sample (see Fig. 4). We cannot say the same for the sample of SDOs
because they reside on larger orbits, making it very difficult to detect or
observe the smaller objects in that population. This selection bias can be
seen in Fig. \ref{Fig4}, where there is only one SDO smaller than 120 km (2002
PN$_{34}$).

\cite{Stansberry2008} showed a tentative correlation (at 2.7$\sigma$)
between albedo and semi-major axis for Centaurs, with all the objects at a
$<$ 20 AU having an albedo $<$ 5\%. Our data do not show this correlation
either for the Centaurs alone, or for the Centaurs plus SDOs (see Table 4).
A tentative positive correlation between albedo and perihelion is also
discussed in \cite{Stansberry2008}, which would suggest that objects at
lower perihelion have a lower albedo. Our results do not confirm this
correlation in either the Centaur or Centaur plus SDOs samples (see Table
4).

Similarly, the correlation between p$_V$ and r(h) previously noticed by
\cite{Stansberry2008} is not confirmed by our results, even when we include
the SDOs (Figure 5, upper panel).  There is no evidence for a correlation between albedo and
inclination or eccentricity. All of this together seems to indicate that the
albedo of Centaurs is not strongly influenced by their dynamical
properties.

\cite{Santos-Sanz2012} reported positive correlations between D and
r(h) and between p$_V$ and r(h) for the SDOs.  The correlation between
D and r(h) exists and is strong for our sample, too, but it is clearer
when we consider both SDO and Centaurs.  We note that the large objects
(D $>$ 120 km) have orbits with  8 $<$ q $<$ 35 AU, while those with
D $<$ 120 km) are restricted to orbits with 5 $<$ q $<$ 20 AU (see
Fig. \ref{Fig6}).  The correlation D vs a is stronger when the SDOs are
included. All of these effects probably result from discovery bias and our
criteria for selecting the Herschel targets.

   \begin{figure}
   \centering
   \includegraphics[width=\columnwidth]{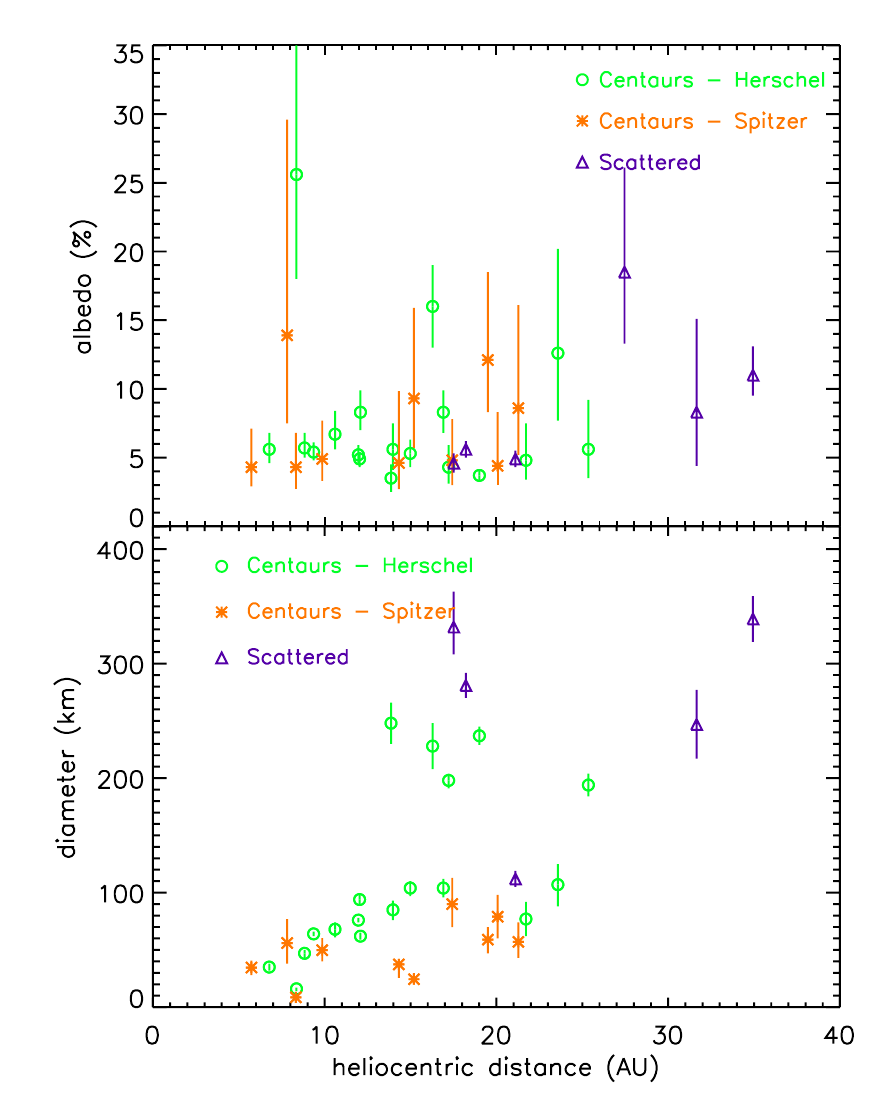}
      \caption{Albedo and diameter versus heliocentric distances. The apparent correlation between heliocentric distance and diameter is probably an observational bias.
              }
         \label{Fig5}
   \end{figure}

   \begin{figure}
   \centering
   \includegraphics[width=\columnwidth]{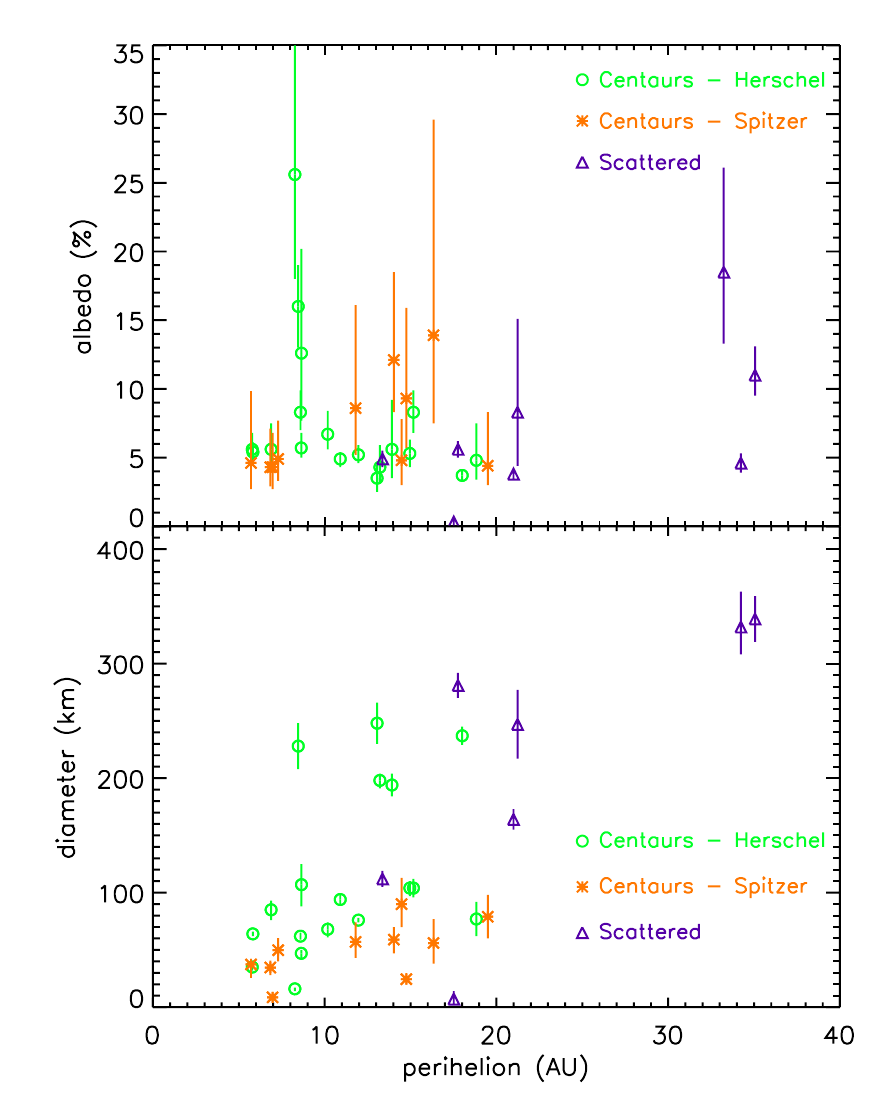}
      \caption{Albedo and diameter versus perihelion distances.
              }
         \label{Fig6}
   \end{figure}

In the plot of albedo vs. spectral slope (fig. 7 top panel), we can clearly see the difference between gray (smaller slope) and red (larger slope) objects. There is no object with slopes from 18 to 28\%/1000 $\AA$, this is just a consequence of the color bimodality in the population. However, other patterns appear in the albedo vs. color representation with most of the gray objects being dark, while the red ones span a broader range of albedos. Even if there are some gray objects with high albedos that had not been included in previous studies, the 74\% of the gray cluster (those with spectral slope $<$ 18\%) have an albedo $<$ 6\%, while all the red objects have an albedo $>$ 5\%, so most of the objects with low albedo are found in the gray group. The gray Centaurs seem to be more clustered in albedo, with a mean value of 5.73$\pm$3.6\% considering Chiron and 29P (the only two objects in this group with pv $>$ 8\%), and 4.8$\pm$1.3\% without them. For the redder objects the albedos can vary in a more continuous way from 5 to 14\% with a mean value of 8.5\%. To test a statistical significance of this result we ran a Kolmogorov-Smirnov test, our null hypothesis being that both samples are derived from the same distribution. We removed the gray Centaur Chiron from the sample, because it was active during the observation. We also discarded 2007 OR$_{10}$ since, as explained before, this red Centaur has a size, albedo, and surface composition very different from the rest of the sample. The KS test rules out the null hypothesis with a p-value of 0.1\%, which means that the albedos of the red and gray groups of centaurs are not drawn from the same distribution.

Radiative transfer models show that mixtures of volatile ice and
non-volatile organics might account for the extremely red surfaces of some TNOs
and the red lobe of the Centaur population \citep{Grundy2009}. For these
red objects with lower perihelia, these materials
could become progressively darker and less red as the ice sublimates
away. However, reality is more complex and additional processes have to be
considered. Cometary activity might trigger episodes of fast sublimation
and the development of mantles of silicates, masking the original ices
and darkening the surface. This is just what \cite{Melita2012} showed,
all the objects in the gray lobe have passed a longer time
in orbits with smaller perihelion. Accordingly, the albedo vs slope distribution
of our sample is compatible with a red lobe of Centaurs richer in ices
and organics with a higher albedo, and a gray lobe covered by surfaces
poor in ice and rich in silicates with lower albedos affected by rapid
sublimation typical of episodes of cometary activity. We can explain
the objects in the gray group with higher albedo as being active
Centaurs that may have some neutral-colored ices exposed on the surface
as a result of that activity or a recent impact.

Another interesting result comes from the comparison of diameter vs
slope. Again we can see the gap between S'=18 and 28 that separates the
gray and red lobe of the Centaurs. The lack of large objects
in the red lobe is evident . All of our objects are smaller than 400 km in diameter
(with the exception of 2007 OR$_{10}$ with 1280 km and the ability to
retain volatiles), which means that all of them are too small to have
retained part of the original inventory of volatiles on their surfaces. We
can draw a horizontal line at 120 km so that all the objects above it
appear in the gray lobe, while objects below this limit appear either
in the red or in the gray lobe. It is known that objects smaller than
100-150 km are fragments of collisions suffered by larger objects; this suggests that collisions might be a factor that affects the different colors of small and large objects.

   \begin{figure}
   \centering
   \includegraphics[width=\columnwidth]{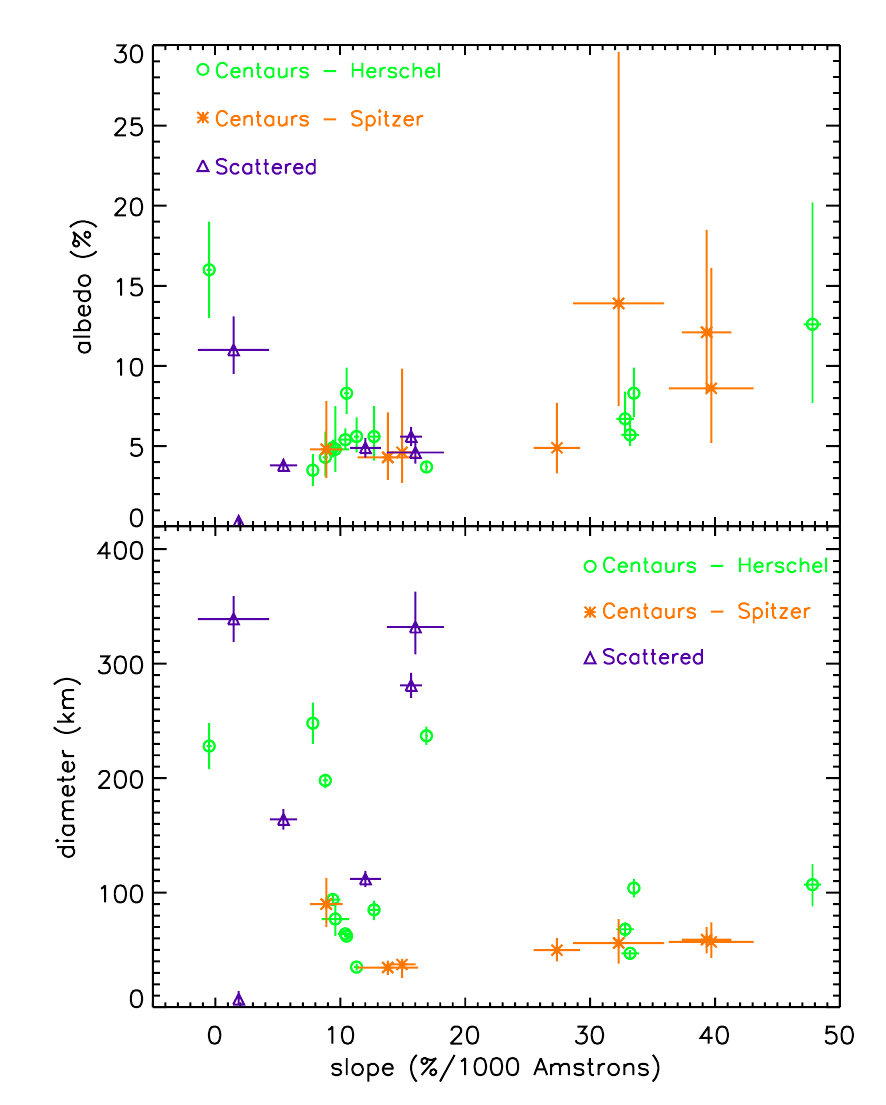}
      \caption{Albedo and diameters versus visible spectral slopes. We can clearly see the gap in the slopes separating the gray and red objects. No red object larger than 120 km is present in the sample.
              }
         \label{Fig7}
   \end{figure}

\cite{Peixinho2012} found that the bimodal color distribution of Centaurs
is a size-related phenomenon, common to both Centaurs and small KBOs, that is, independent of dynamical classification. We analyzed the smallest
object sample and no strong correlations are present when we consider
only the smaller Centaurs.

A low-significance 2.2$\sigma$ medium correlation between albedo and
spectral slope is, nonetheless, detected.

\section{Discussion}

Centaurs are smaller than most of the observed TNOs, and some of them
were active recently or are currently active.  This cometary activity
can result in inhomogeneous surfaces, where two, or more different
terrains are possible. In the near future, thermo-physical models should
be applied in the Centaur population. Stellar occultation data
are also valuable and can constrain diameters and albedos.

While \cite{Santos-Sanz2012} found a correlation between albedo
and diameter in a sample of SDOs,  one of the main results here
is the absence of an albedo/diameter correlation in a larger
sample. We analyzed here only a limited sample of Centaurs, only those 
with a radiometric diameter/albedo determination. But other analyse can
be made using the H magnitude as a simile of diameter and assuming an
albedo \citep{Peixinho2012}. Care must be taken in choosing the albedo
to transform H magnitude to diameters. Here we show that for the larger
Centaurs, the mean albedo is 4.2\% (four objects, with the exception of the
active Chiron), while the smaller Centaurs have a larger dispersion
on albedo, from 4\% to 15\% (with the exception of 2005 UJ$_{438}$).

Our results show an excellent concordance with those of
\cite{Bauer2013} in both diameters and albedos. However,  our errors
are smaller which results in a more precise estimate for
individual objects. The error in the mean of any variable is the standard deviation. Objects whose differences are not within the
error bar can be explained by a combination of different effects:
a different value of the absolute magnitude, the lack of observations
on some of the bands, or the different observing conditions. Our results
are also equivalent within the 1$\sigma$ error bar when we studied the
mean albedo of the population. \cite{Bauer2013} found a mean albedo of
8$\pm$4\% while we found a similar value of 6.9$\pm$4.8\% for the
Centaurs and a 6.7$\pm$4.8\% for the whole sample.  Another interesting
result is the difference in albedos within the Centaur population. The mean albedos of our red and gray objects are 8.5$\pm$4.9\%
and 5.6$\pm$1.2\%, respectively. A similar result was also observed by
\cite{Bauer2013}, with a mean albedo of 12$\pm$5\% for the redder objects
and of 6$\pm$2\% for the gray ones.  Our results also show that the size
distribution of the Centaurs in both groups is different, there are
no large objects in the red lobe. 

For the origin of the Centaurs we can add that the Centaur region
is being resupplied, but the exact source region is unknown. They may
be coming from the dynamical class of scattered-disk TNOs or from an yet
unknown source \citep{Volk2008}, for example as from the inner Oort cloud
\citep{Emelyanenko2005}, and there may be an exchange between the Centaur
region and the Trojans of Jupiter and Neptune. Comparing the color distribution of Jupiter Trojans
 with those of other bodies, \cite{Fornasier2007}
found that the Jupiter Trojans and neutral/gray Centaurs have fairly similar
mean colors, but different color distributions. If there are several source
regions, what is the contribution from each of them?

\cite{Disisto2009} found that the vast majority of escaped Plutinos have
encounters with Neptune and that this planet governs their dynamical
evolution. When a Plutino escapes from the resonance, it is transferred to
either the scattered-disk zone (q$>$30 AU) or the Centaur zone (q$<$30 AU),
but eventually switches from Centaurs to SDOs or vice versa because of the
dynamical influence of Neptune. The escaped Plutinos would have a mean
lifetime in the Centaur zone of 108 Myr and their contribution to the
Centaur population would be somewhat smaller than 6\% of the total Centaur
population. In this way, escaped Plutinos would be a secondary source of
Centaurs. This is consistent with the
 findings of \cite{Mommert2012}, who found no correlation
 between the diameters and albedos of Plutinos.

Some Centaurs (10\%) show comet-like activity even though they are
far away from the Sun. At distances outside the water zone, that is 5-6
AU, such activity cannot be explained by direct water-ice sublimation
alone \citep{Meech2004}. This suggests that mass loss is driven by a
process other than the sublimation of water ice. There have been some
attempts to explain this by the crystallization of amorphous ice via
the release of trapped gas (\cite{Capria2000, Notesco2003}), or CO$_2$
ice \citep{Jewitt2009}. On the other hand, some Centaurs remain inactive
even though their perihelia are small.

One scenario could be that the entire Centaur population is a mixture from
different sources: Jupiter Trojans, Plutinos, SDOs, and the inner Oort
cloud. Moreover, we are ignorant of the size of the object that became a
Centaur. Smaller objects might be injected from the original source or be pieces of larger objects that were disrupted. This agrees with
\cite{Peixinho2012} who found that both small Centaurs and small TNOs have
a color bimodality.

In some respects the Centaurs resemble the near Earth objects (NEOs)
 of the trans-Neptunian region. Like the NEOs, they are dynamically unstable
 and short-lived, and they are thought to sample different populations of the
 TN belt. Given their diverse sources, why look for correlations between their
 properties at all? One reason is that this is a test of the idea that they are drawn
 from diverse parent populations: discovering a strong correlation in the
 Centaurs that would match a correlation in another TNO population would call that
 hypothesis into question. Another reason is that individual Centaurs experience a different thermal evolution, because their orbital parameters, and their
 physical characteristics might provide clues about the nature of that evolution. As \cite{Melita2012} showed, gray/red colors are related with
the time the objects spent near the Sun and are not not related with the actual perihelion
distance. Thermal evolution (de-volatilization) of Centaurs might also lead to measurable changes in their size. This might be an interesting area
 for future modeling, and might provide new insight into the sizes we present here.\\

\section{Summary and conclusions}

Our main findings are summarized as follows:
\begin{itemize}
\item We presented 16 diameters and albedos (and the beaming factors where possible) of Centaurs observed with Herschel-PACS. We merged
our results with 12 other radiometric results from the literature
to statistically analyze the distribution of sizes
and albedos. We also used the albedos and diameters of eight
SDOs obtained with Herschel. The final sample consisted of 28 Centaurs and 8 SDOs. 

\item Most of the Centaurs in our sample are small objects (82\% are
smaller than 120km) and the distribution of sizes is bimodal with a
lack of objects with sizes between 120 and 190 km.

\item Most of the Centaurs in our sample are low-albedo objects (61\%
have a p$_v$ $<$ 6\%), the remaining objects show an albedo between
6 and 16\% (with the exception of 2005 UJ$_{438}$).

\item There is no correlation between diameter and albedo. The mean albedo for our small and large Centaurs is 7.0$\pm$4.0\% and 6.2$\pm$5.0\%, respectively.

\item The albedo of the Centaurs is not correlated with their orbital
parameters. No
correlation is found between the albedo and the orbital parameters
(a,e,i,q) and r(h). The same is true for the diameter we found no correlation. 

\item When we compared albedo and diameter
with the surface properties (spectral slope) we clearly saw the
typical bimodal distribution, with a gray lobe formed by objects
with S'$<$18\% and a mean albedo of 5.6\%; and a red lobe formed by those with a
spectral slope larger than 28\% and a mean albedo of 8.5\%. Moreover,
the color distribution of the small and large objects is different. Our
data clearly showed that all the large Centaurs are gray, while the small
ones can be found in either the red or gray group. Color bimodality
is only present in small Centaurs.

\end{itemize}

\begin{acknowledgements}
RD acknowledges the support of MINECO for his Ramon y Cajal Contract. N.P.A's work was supported by MINECO through a fellowship of the Juan de la Cierva program. A.P. acknowledges grant LP2012-31. N.P.A. and P. S-S. acknowledge support by contract AYA2011-30106-C02-01. The work of C.K. has been supported by the Hungarian Space Office and the European Space Agency through contract PECS-98073, the K-104607 grant of the Hungarian Research Fund (OTKA), and the Bolyai Research Fellowship of the Hungarian Academy of Sciences. EV acknowledges the support of the German \emph{DLR} project number 50 OR 1108. This work is based [in part] on observations made with the Spitzer Space Telescope, which is operated by the Jet Propulsion Laboratory, California Institute of Technology under a contract with NASA. NP acknowledges funding by the Gemini-Conicyt Fund, allocated to project N\textsuperscript{\underline{o}} 32120036.  
\end{acknowledgements}

\bibliographystyle{aa}
\bibliography{centaurs}



\begin{landscape}
\begin{table}
\caption{Centaur orbital and physical data. All orbital parameters are from the Minor Planet Center (MPC). The table lists the semi-major axis, the eccentricity, and the inclination of each Centaur as well as the rotational period in hours (errors are presented when available in literature) and light-curve amplitude in magnitudes. The last 3 columns are the taxonomic classification, the visible spectral slope and information on the surface composition (see the description of each Centaur in the text for references). Also in last column there is an indication if the object overlaps with the WISE sample, indicated with (Wise).  This table also presents the data for Chiron and Chariklo published in \cite{Fornasier2013}. }  
\label{table-1}
\begin{tabular}{l r r r r c c c c c }     
\hline\hline
\noalign{\smallskip}
\multicolumn{10}{c}{Herschel Centaurs} \\
\hline
Name & a & q & e & i & Rot.           &  LC      & Tax. & Spec. Slope & Ices \\
          &   &     &    &   & Period [hr]       & Ampl.  &          & \%/1000 $\AA$ &  \\
\hline
(95626) 2002 GZ$_{32}$  & 22.99     & 18.00     & 0.217 & 15.030    &  5.80     & 0.08      & BR  & 16.9$\pm$0.1 & Water? (Wise)\\
(250112) 2002 KY$_{14}$ & 12.62     & 8.63  & 0.316 & 19.456    & 3.56  & 0.13      & RR & 33.2$\pm$ 0.7 & Water? maybe (Wise) \\
(120061) 2003 CO$_1$    & 20.68     & 10.91     & 0.472 & 19.763    & 10        & 0.07 - 0.10   & BR  & 9.4$\pm$0.6 & Water? maybe (Wise)\\
(136204) 2003 WL$_7$    & 20.27 & 14.95     & 0.262 & 11.163    & 8.24  & 0.05      & BB  & \ldots & (Wise)\\
2005 RO$_{43}$          & 29.00     & 13. 93    & 0.520 & 35.415    &  \ldots   & \ldots        & \ldots  & \ldots & \\
(145486) 2005 UJ$_{438}$     & 17.70    & 8.26  & 0.533 & 3.783              & 8.32     & 0.13      & RR, IR & \ldots  & (Wise) \\
(248835) 2006 SX$_{368}$     & 22.25    & 11.96     & 0.462 & 36.287    &  \ldots   & \ldots        & BR  & \ldots  & (Wise)\\
(281371) 2008 FC$_{76}$     & 14.79     & 10.18     & 0.312 & 27.105    &  \ldots   & \ldots        & RR & 32.8$\pm$0.7 & no Water (Wise)\\
(55576) Amycus          & 24.95     & 15.18     & 0.392 & 12.345    & 9.76  & 0.16      & RR  & 33.5$\pm$0.1 & Water (Wise) \\
(8405) Asbolus          & 18.10     & 6.88  & 0.620 & 17.610    & 8.935     & 0.32 - 0.55   & BR  & 12.7$\pm$0.4  & no ice \\
(54598) Bienor          & 16.56     & 13.24     & 0.201 & 20.731    & 9.14  & 0.34 - 0.75   & BR  & 8.8$\pm$0.2  & Water (Wise) \\
(10199) Chariklo             & 15.74    & 13.06     & 0.170 & 23.400    & \ldots    & $<$0.1        & BR  & 7.8$\pm$0.1 & Water (Wise) \\
(2060) Chiron           & 13.67     & 8.49  & 0.379 & 6.927              & 5.918    & 0.04 - 0.09   & BB    & -0.5$\pm$0.2 & Water (Wise) \\
(60558) Echeclus                & 10.71 & 5.82  & 0.457 & 4.342              & 26.802   & \ldots        & BR  & 10.4$\pm$0.6 & any (Wise) \\
(10370) Hylonome        & 24.96     & 18.85     & 0.245 & 4.150              &  \ldots  & $<$0.04            & BR  & 9.6$\pm$1.1 &   \\
(52872) Okyrhoe               & 8.34    & 5.79  & 0.306 & 15.660    & 9.72  & 0.07 - 04     & BR  & 11.3$\pm$0.2 & Water, maybe. (Wise) \\
(5145) Pholus           & 20.28     & 8.65  & 0.573 & 24.705    & 9.98  & 0.15 - 0.60   & RR  & 47.8$\pm$0.7 & Water, Methanol\\
(32532) Thereus             & 10.67     & 8.58  & 0.196 & 20.329    & 8.335     & 0.18 - 0.38   & BR  & 10.5$\pm$0.2 & Water. (Wise)\\
\hline
\hline
\noalign{\smallskip}
\multicolumn{10}{c}{Spitzer un-published Centaurs} \\
\hline
\hline
(119315) 2001 SQ$_{73}$     & 17.52 & 14.47 & 0.174 & 17.4 & \ldots & \ldots                      & \ldots  & 8.875$\pm$1.301   &     \\
(119976) 2002 VR$_{130}$   & 24.11 & 14.75 & 0.388 &  3.5 & \ldots & \ldots                      &  \ldots         & \ldots             &   \\
2004 QQ$_{26}$                  & 22.88 & 19.50 & 0.148 & 21.5 & \ldots & \ldots                     &  \ldots         &  \ldots              &    \\
2000 GM$_{137}$             &  7.90 &  6.97 & 0.118 & 15.8 & \ldots & \ldots                     &    \ldots          & \ldots               &   \\
\hline
\multicolumn{10}{c}{Spitzer re-analyzed targets} \\
\hline
29P                        & 6.00 & 50.73  & 0.042  & 9.37  & 288$\pm$24           &     \ldots   & \ldots &          14.94$\pm$1.09                 & (Wise)   \\
(7066) Nessus     & 24.48 & 11.80 & 0.518 & 15.6 & \ldots & \ldots                             &\ldots & 39.7 $\pm$ 3.4 &   \\
(31824) Elatus    & 11.80 &  7.29 & 0.382 &  5.2 & $13.41\pm 0.04$ & $0.102\pm 0.005$   & RR  & 27.34$\pm$1.86   & Water. (Wise)  \\
(52975) Cyllarus    & 26.30 & 16.34 & 0.379 & 12.6 & \ldots & \ldots                             & \ldots  & 32.28$\pm$3.65  &   \\
(83982) Crantor     & 19.35 & 14.03 & 0.275 & 12.8 & \ldots & \ldots                             & RR  & 39.22 $\pm$1.99 & Water  \\
(63252) 2001 BL$_{41}$   &  9.72 &  6.83 & 0.297 & 12.5 & \ldots & \ldots                             & BR  & 13.8$\pm$2.42  & None?   \\
\end{tabular}
\end{table}
\end{landscape}


\begin{table*}
\caption{Individual observations of the sample of 16 Centaurs observed by Herschel-PACS and 4 Centaurs observed by Spitzer-MIPS. The results on these last 4 objects were not published before. The first line for each target  contains the 4 subsequent measurements (twice the blue band, twice the green band, and four times the red band) for the first visit, the second line contains the measurements of the second visit, after the target has moved by a few beams. The identification number of the observation is given, followed by the duration and mid-time of the exposure. We also present the heliocentric distance (r), geocentric distance ($\Delta$), and phase angle ($\alpha$) of the observation. }
\label{table-2}
\centering
\begin{tabular}{lcccccc}
\hline
Target  & OBSIDs & Duration & Mid-time  & $r$   & $\Delta$  & $\alpha$  \\
    &    & (sec.)   &       & (AU)  & (AU)      & ($^{\circ}$)  \\
\hline
\hline
(95626)  2002 GZ$_{\mathrm 32}$ & 1342202937-40 &  568 & 2010-08-12 17:19:21 & $19.006$ & $19.232$ & $3.0$ \\
            & 1342202967-70 &  568 & 2010-08-13 01:39:22 & & & \\
(250112)  2002 KY$_{\mathrm 14}$    & 1342211112-15 &  850 & 2010-12-13 16:37:30 & $ 8.824$ & $ 8.547$ & $6.2$ \\
            & 1342211144-47 &  850 & 2010-12-13 22:26:46 & & & \\
(120061)  2003 CO$_{\mathrm 1}$ & 1342202345-48 &  568 & 2010-08-09 18:23:36 & $12.032$ & $11.842$ & $4.8$ \\
            & 1342202361-64 &  568 & 2010-08-10 07:15:55 & & & \\
(136204) 2003 WL$_{\mathrm 7}$  & 1342191941-44 &  850 & 2010-03-10 01:03:04 & $14.980$ & $15.217$ & $3.7$ \\
            & 1342191966-69 &  850 & 2010-03-10 09:42:50 & & & \\
2005 RO$_{\mathrm 43}$  & 1342212848-51 &  850 & 2011-01-18 06:55:29 & $25.352$ & $25.016$ & $2.1$ \\
            & 1342213115-18 &  850 & 2011-01-19 07:03:23 & & & \\
(145486)  2005 UJ$_{\mathrm 438}$ & 1342218768-71 & 1414 & 2011-04-18 01:37:12 & $ 8.356$ & $ 8.272$ & $6.9$ \\
            & 1342218784-87 & 1414 & 2011-04-18 07:43:45 & & & \\
(248835) 2006 SX$_{\mathrm 368}$ & 1342196759-62 &  850 & 2010-05-20 13:14:29 & $11.961$ & $12.246$ & $4.6$ \\
            & 1342196771-74 &  850 & 2010-05-20 18:22:09 & & & \\
(281371)  2008 FC$_{\mathrm 76}$    & 1342222926-29 &  568 & 2011-06-22 08:34:29 & $10.599$ & $10.574$ & $5.5$ \\
            & 1342222933-36 &  568 & 2011-06-22 13:37:26 & & & \\
Amycus      & 1342202341-44 &  850 & 2010-08-09 17:28:22 & $16.906$ & $16.626$ & $3.3$ \\
            & 1342202367-70 &  850 & 2010-08-10 08:38:51 & & & \\
Asbolus         & 1342190921-24 &  548 & 2010-02-22 00:35:35 & $13.973$ & $14.316$ & $3.8$ \\
            & 1342190937-40 &  548 & 2010-02-22 06:40:11 & & & \\
Bienor      & 1342213252-55 &  568 & 2011-01-24 12:52:20 & $17.216$ & $17.553$ & $3.0$ \\
            & 1342213274-77 &  568 & 2011-01-24 22:00:27 & & & \\
Echeclus        & 1342201153-56 &  568 & 2010-07-23 18:09:47 & $ 9.353$ & $ 9.008$ & $6.0$ \\
            & 1342201194-97 &  568 & 2010-07-25 13:33:02 & & & \\
Hylonome        & 1342215386-89 & 1132 & 2011-03-07 00:25:13 & $21.717$ & $21.709$ & $2.6$ \\
            & 1342215607-10 & 1132 & 2011-03-08 00:33:05 & & & \\
Okyrhoe         & 1342202865-68 &  568 & 2010-08-11 17:49:38 & $ 6.771$ & $ 7.044$ & $8.1$ \\
            & 1342202893-96 &  568 & 2010-08-12 00:23:43 & & & \\
Pholus      & 1342205148-51 &  850 & 2010-09-26 22:58:03 & $23.576$ & $23.915$ & $2.3$ \\
            & 1342205153-56 &  850 & 2010-09-27 08:15:50 & & & \\
Thereus         & 1342216137-40 &  568 & 2011-03-06 03:09:48 & $12.082$ & $12.190$ & $4.7$ \\
            & 1342216150-53 &  568 & 2011-03-06 10:32:29 & & & \\
\hline
\noalign{\smallskip}
\multicolumn{7}{c}{Spitzer sample unpublished Centaurs} \\
\hline
Target  & AORKEY & Duration & Mid-time  & $r$   & $\Delta$  & $\alpha$  \\
    &    & (sec)   &        & (AU)  & (AU)      & ($^{\circ}$)  \\
\hline
\hline
(119315) 2001 SQ$_{73}$ & 26025216, 26029824  & 872 & 2009-03-24 10:14:21 & 17.427  & 17.354  & 3.3  \\
(119976) 2002 VR$_{130}$    & 26026752, 26031360  & 872 & 2009-03-25 07:23:28 & 15.199  & 14.982  & 3.7 \\
2004 QQ$_{26}$          & 26027520, 26032128  & 872 & 2009-03-24 05:45:20 & 20.070  & 20.030  & 2.8  \\
2000 GM$_{137}$              & 26026496, 26031104  & 872 & 2008-11-27 03:50:33 &  8.321  &  7.775  & 6.1  \\
\hline
\end{tabular}
\end{table*}


\begin{table*}
\caption{Individual absolute magnitudes (H$_{mag}$) and Herschel flux densities ($F_{70}$, $F_{100}$ and $F_{160}$), diameter (D), albedo (p$_v$), and beaming parameter ($\eta$) of the Centaur sample. Color-corrected Spitzer
flux densities ($F_{24}$ and $F_{71}$) of previously published targets and four unpublished targets. The object 29P is excluded from this table because we used the values published in \cite{Stansberry2008}. All the used  H$_{mag}$ also consider the amplitude of the light-curve adding quadratically the 88\% of half the amplitude to each uncertainty (when amplitude is unknown we assumed an amplitude of 0.2 mag -\citep{Vilenius2012}-. For some of the objects this variation is significant, about 0.3-0.4 mag (e.g.  Pholus and Asbolus). The MIPS and PACS data have been taken at very different epochs, hence have substantially different observing geometries.} 
\label{table-3}
\centering
\begin{tabular}{lccccccccc}
\hline
\noalign{\smallskip}
\multicolumn{10}{c}{Herschel sample} \\
\noalign{\smallskip}
\hline
\noalign{\smallskip}
Target                          & $H_{mag}$                       & $F_{24}$                  & $F_{71}$                         & $F_{70}$               & $F_{100}$            & $F_{160}$             &   $D$                       & $p_v$                       & $\eta$        \\
                                    & (mag)                        &  (mJy)                     &  (mJy)                          & (mJy)                      & (mJy)                   & (mJy)                    & (km)                      & \%                            &                   \\
\hline
\noalign{\smallskip}
2002 GZ$_{32}$  & $7.37\pm 0.10 (a)$  & $9.88\pm 0.31$  & $44.49\pm 2.91 $            & $54.48\pm1.80 $ & $41.16\pm 1.94 $ &  $20.26\pm2.17$ & 237$^{+8}_{-8}$   & 3.7$^{+0.4}_{-0.4}$ & 0.97$^{+0.05}_{-0.07}$     \\
\noalign{\smallskip}
2002 KY$_{14}$  & $10.37\pm0.07 (b)$ &         .....                        &            .....                       & $23.43\pm 1.20 $ & $14.81\pm 0.92 $ & $ 6.76\pm1.29$  & 47$^{+3}_{-4}$     & 5.7$^{+1.1}_{-0.7}$ &  1.20$^{+0.35}_{-0.35}$   \\
\noalign{\smallskip}
2003 CO$_{1}$   & $9.07\pm 0.05 (c)$      & $21.12\pm 0.68$ &$34.02\pm 3.60$   & $34.51\pm 1.45 $ & $23.73\pm 2.11 $  & $11.28\pm1.63$ & 94$^{+5}_{-5}$    & 4.9$^{+0.5}_{-0.6}$ & 1.23$^{+0.12}_{-0.11}$ \\
\noalign{\smallskip}
2003 WL$_{7}$   &$8.75\pm 0.16 (d)$       & $7.08\pm 0.21$  & $20.95\pm 1.54 $  & $22.18\pm 1.30$  & $15.07\pm 1.16 $   & $ < 3.04     $    & 105$^{+6}_{-7}$  & 5.3$^{+1.0}_{-1.0}$ & 1.02$^{+0.07}_{-0.05}$ \\
\noalign{\smallskip}
2005 RO$_{43}$  & $7.34\pm 0.51 (e)$  & $ 0.81\pm 0.03$ &     .....                               & $11.30\pm 0.93$  & $12.69\pm 1.24 $   & $ 5.77 \pm2.48$ & 194$^{+10}_{-10}$  & 5.6$^{+3.6}_{-2.1}$ & 1.12$^{+0.05}_{-0.08}$ \\
\noalign{\smallskip}
2005 UJ$_{438}$    & $11.14\pm 0.32 (f)$    & $ 8.05\pm 0.24$  & $4.71\pm 0.96$    & $ 5.02\pm 0.72 $  & $ 3.38\pm 0.89 $    & $2.78\pm1.56$    & 16$^{+1}_{-2}$     & 25.6$^{+9.7}_{-7.6}$ & 0.34$^{+0.09}_{-0.08}$    \\
\noalign{\smallskip}
2006 SX$_{368}$    &$9.45\pm 0.11 (g)$   &$15.09\pm 0.48$ & $26.65\pm 2.40$ & $25.00\pm 1.12$  & $15.63\pm 1.21$    & $10.71\pm1.94$    &  76$^{+2}_{-2}$    & 5.2$^{+0.7}_{-0.6}$ & 0.87$^{+0.04}_{-0.06}$  \\
\noalign{\smallskip}
2008 FC$_{76}$  & $9.44\pm 0.10 (h)$   &       .....                         &       .....                             & $26.79\pm1.43 $  & $15.58\pm 1.72 $   & $ 5.00\pm1.65 $ & 68$^{+6}_{-7} $    & 6.7$^{+1.7}_{-1.1}$ & 1.20$^{+0.35}_{-0.35}$    \\
\noalign{\smallskip}
Amycus                   & $8.27\pm 0.11 (i)$   &$ 6.19\pm 0.20$ & $14.74\pm 1.93$   & $17.02\pm 0.82$   & $ 10.31\pm 0.91$   & $ 6.53\pm3.16$  & 104$^{+8}_{-8}$   & 8.3$^{+1.6}_{-1.5}$ & 1.00$^{+0.12}_{-0.13}$ \\
\noalign{\smallskip}
Asbolus(7)              &$9.13\pm 0.25 (j)$   & $73.52\pm 2.24$  & $85.70\pm 9.56$            & $  8.51\pm 1.62 $   & $ < 2.07 $  & $ < 2.99 $      & 85$^{+8}_{-9}$     & 5.6$^{+1.9}_{-1.5}$ & 0.97$^{+0.14}_{-0.18}$ \\
\noalign{\smallskip}
Bienor              &$7.57\pm 0.34 (k)$   & $3.51\pm 0.11$  & $28.81\pm 4.77$   & $37.51\pm 1.42  $ & $26.67\pm 1.36 $ & $14.45\pm4.92 $& 198$^{+6}_{-7}$  & 4.3$^{+1.6}_{-1.2}$ & 1.58$^{+0.07}_{-0.07}$ \\
\noalign{\smallskip}
Echeclus(1)                 &  $ 9.78\pm 0.14 (l)$               & $4.91\pm 0.15$      & $16.36\pm 4.86$      & 42.15$\pm$ 1.50 &   26.93$\pm$ 2.17& $16.55\pm2.11$ & 65$^{+2}_{-2}$    &   5.3 $^{+0.7}_{-0.7}$    & 0.87 $^{+0.04}_{-0.04}$      \\
Echeclus(2)                 & $ 9.78\pm 0.14 (l)$              & $20.15\pm 0.61$    &        .....                        & $42.15\pm1.50$  & $26.93\pm2.17$ &  $16.55\pm2.11$   &63 $^{+2}_{-2}$   &   5.5 $^{+0.9}_{-0.6}$    & 0.81 $^{+0.04}_{-0.05}$  \\
{\bf Echeclus}          &  &                       &                                      &                          &                   &                      &  {\bf 64.6$^{+1.6}_{-1.6}$}    &{\bf 5.2$^{+0.70}_{-0.71}$}   & {\bf 0.86$^{+0.036}_{-0.037}$} \\
\noalign{\smallskip}
Hylonome(3)                 & $ 9.51\pm 0.08 (m)$  &  $0.51\pm 0.08$         &   $ < 4.798 $          &$ 2.76\pm 0.96 $ & $2.06\pm 1.09$    &  $ 4.06\pm2.45$  & 77$^{+15}_{-16}$   & $4.9^{+2.7}_{-1.7}$ & $1.32^{+ 0.34}_{-0.32}$\\
Hylonome(4)                 & $ 9.51\pm 0.08 (m)$  &  $0.44\pm 0.03$         &       .....                        & $ 2.76\pm 0.96 $    & $2.06\pm 1.09$    &  $ 4.06\pm2.45$  &  77$^{+17}_{-15}$   & $4.7^{+2.8}_{-1.4}$ & $1.35^{+0.31}_{-0.30}$\\
{\bf Hylonome }         &                        &   &    &                     &                   &                       & {\bf 74$^{+16}_{-16}$} & {\bf 5.1$^{+3.0}_{-1.7}$} & {\bf 1.29$^{+0.31}_{-0.31}$}\\
\noalign{\smallskip}
Okyrhoe                 &$11.07\pm 0.10 (n)$      & $28.89\pm 0.89$ &$37.53\pm 5.07$   & $30.34\pm 1.80 $   & $14.82\pm 2.50 $ & $10.02\pm3.64$   & 35$^{+3}_{-3}$    & 5.6$^{+1.2}_{-1.0}$ & 0.71$^{+0.12}_{-0.13}$ \\
\noalign{\smallskip}
{\bf Pholus(5)}               & $ 7.68\pm 0.28 (o)$    &  $3.13\pm 0.13$  & $ 15.63\pm7.11$                                 & $ 4.83\pm 0.99 $    &  $ 3.21\pm1.86$    & $ <3.91$ & {\bf 99$^{+15}_{-14}$ } &  {\bf  15.5$^{+7.6}_{-4.9}$  }  &{\bf  0.77$^{+0.16}_{-0.16}$} \\
Pholus(6)               & $ 7.68\pm 0.28 (o)$    &   $0.95\pm 0.10$ & $  < 5.220$                                    & $ 4.83\pm 0.99 $ &  $ 3.21\pm1.86$    & $ <3.91$ & $119^{+18}_{-19}$  &  $11.0^{+5.7}_{-3.6}$     & $1.48^{+0.30}_{-0.28}$\\
\noalign{\smallskip}
Thereus(8)              &$9.40\pm 0.16 (p)$   &$23.60\pm 0.73$ & $43.90\pm5.44$   & $17.39\pm1.22$  & $9.68\pm1.08 $   & $4.46\pm2.05$ & 62$^{+3}_{-3}$  & 8.3$^{+1.6}_{-1.3}$ & 0.87$^{+0.08}_{-0.08}$ \\
\hline
\hline
\noalign{\smallskip}
\multicolumn{10}{c}{Spitzer sample re-ananlyzed targets} \\
\hline
\noalign{\smallskip}
Target                                    & $H_{mag}$                       & $F_{24}$                    & $F_{71}$                         & $F_{70}$               & $F_{100}$            & $F_{160}$             &   $D$                         & $p_v$                        & $\eta$        \\
                                              & (mag)                        &  (mJy)                     &  (mJy)                          & (mJy)                      & (mJy)                   & (mJy)                    & (km)                       & \%                              &                   \\
\hline
\noalign{\smallskip}
Nessus                              &   $9.51\pm 0.22$ (s)                                & 0.38$\pm$ 0.09    &         .....              &         .....                       &     .....                     &      .....                        & 57$^{+17}_{-14}$ & 8.6$^{+7.5}_{-3.4}$  & 1.20$^{+0.35}_{-0.35}$ \\
\noalign{\smallskip}
 Elatus/2004                     &                                  & 8.59$\pm$ 0.12     & $<$ 5.97                    &      .....                          &       .....                    &        .....                     &  & & \\
\noalign{\smallskip}
 Elatus/2005                    &  $10.40\pm 0.09$ (t) & 6.05$\pm$ 0.20  & 10.68$\pm$4.30          &           .....                      &    .....                       &      .....                       & 49.8$^{+10.4}_{-9.8}$ & 4.9$^{+2.8}_{-1.6}$ & 1.20$^{+0.35}_{-0.35}$\\
\noalign{\smallskip}
Cyllarus                        & $9.02\pm 0.15$ (v)  & 0.21$\pm$ 0.11 &        .....                     &           .....                      &        .....                  &        .....                      & 56$^{+21}_{-18}$ & 13.9$^{+15.7}_{-6.4}$ & 1.20$^{+0.35}_{-0.35}$\\
\noalign{\smallskip}
2001 BL$_{41}$              & $11.34\pm 0.21$ (w) & 4.90$\pm$ 0.15 &   .....            &          .....                       &            .....              &     .....                        & 34.6$^{-6.1}_{+6.6}$   &  4.3$^{+2.8}_{-1.4}$ & 1.20$^{+0.35}_{-0.35}$ \\
\noalign{\smallskip}
 Crantor                         &  $9.03\pm 0.16$ (w) & 2.26$\pm$0.08 & $<$ 6.31                    &         .....                        &     .....                     &       .....                       & 59$^{+11}_{-12}$ & 12.1$^{+6.4}_{-3.8}$ & 1.20$^{+0.35}_{-0.35}$ \\
\noalign{\smallskip}
2001 SQ$_{73}$             &  $9.15\pm 0.11$ (q) & 1.81$\pm$ 0.06 &   .....            &            .....                     &     .....                     &          .....                    & 90$^{+23}_{-20}$& 4.8$^{+3.0}_{-1.8}$& 1.20$^{+0.35}_{-0.35}$ \\
\noalign{\smallskip}
2002 VR$_{130}$           & $11.26\pm 0.39$ (r) & 0.27$\pm$ 0.02 &    .....            &           .....                      &           .....                &    .....                         & 24.4$^{+5.4}_{-4.6}$ & 9.3$^{+6.6}_{-3.6}$ & 1.20$^{+0.35}_{-0.35}$ \\
\noalign{\smallskip}
2004 QQ$_{26}$                 &  $9.53\pm 0.36$ (r) & 0.63$\pm$ 0.02 &     .....           &           .....                      &       .....                    &      .....                       & 79$\pm$19 & 4.4$^{+3.9}_{-1.4}$ & 1.20$^{+0.35}_{-0.35}$ \\
\noalign{\smallskip}
2000 GM$_{137}$           &  $14.36\pm 0.38$ (r) & 0.75$\pm$ 0.03 &   .....            &           .....                      &    .....                       &     .....                         & 8.6$\pm$1.5 & 4.3$^{+2.6}_{-1.6}$ & 1.20$^{+0.35}_{-0.35}$ \\
\noalign{\smallskip}
\hline
\hline
\end{tabular}
\tablebib{(1) MIPS AORKEY 8808960 (2) MIPS AORKEY 26034432 (3) MIPS AORKEY 9038080 (4) MIPS AORKEY 12659968 (5) MIPS AORKEY 9040896 (6) MIPS AORKEY 12661760 (7) MIPS AORKEY 12660480 (8) MIPS AORKEY 9044480 (a) \cite{Rabinowitz2007, Romanishin2005, Doressoundiram2005}; (b) \cite{Perna2010}; (c) \cite{Perna2013}; (d) \cite{Perna2013}; (e) Herschel Database, Delsanti priv. communication 2013; (f) \cite{Perna2013}; (g) Perna \& Dotto (unpublished); (h) \cite{Perna2010} and Perna \& Dotto (unpublished); (i) \cite{Perna2010}; (j) \cite{Rabinowitz2007, Romanishin2005}; (k) \cite{Rabinowitz2007, Romanishin2005, Doressoundiram2005}; (l) Delsanti \& Vilenius (priv. comm. 2013); (m) \cite{Romanishin2005, Doressoundiram2005}; (n) \cite{Romanishin2005, Perna2010, Doressoundiram2005}; (o) \cite{Romanishin2005, Perna2010}; (p) \cite{Rabinowitz2007, Romanishin2005} (q) \cite{Tegler2003, Santos-Sanz2009}; (r) from MPC R-band data (7--24 data points/target) using average V-R=$0.57\pm 0.13$ for Centaurs from MBOSS-2 \cite{Hainaut2012} and average $\beta=0.09\pm 0.04$ from \cite{Perna2013}; (s) Photometric data (N=1) from \cite{Romanishin1997} with new $\beta_\mathrm{V}$ fit from MPC/Steward observatory data (N=34); (t) \cite{Belskaya2003} and V-R color; (v) \cite{Tegler2003, Delsanti2001, Doressoundiram2002, Boehnhardt2001} new slope coefficient fit $\beta=0.141\pm 0.073$; (w) \cite{Bauer2003, Tegler2003, Doressoundiram2005, Fornasier2004, deMeo2009}, default $\beta=0.09\pm 0.04$; (w) \cite{Tegler2003}, default $\beta=0.09\pm 0.04$
}
\end{table*}


\begin{table*}
\begin{center}
\caption{Correlation results on the different Centaur samples. Strongest correlations are listed in boldface. }
\label{tab:cent_correlations}
\renewcommand{\arraystretch}{1.2}
\begin{tabular}[]{lrrcr}
\hline
\multicolumn{5}{c}{Only Centaurs } \\
\hline \hline
 Variables  & Number of   & Correlation  & Significance & Confidence \\
                    &  data points & coefficient     &                        & level ($\sigma$) \\
\hline
${ D }$ {\it vs.} ${ p_V }$ & $  28 $ & $ -0.21 ^{+ 0.27 }_{- 0.25 }$ & $ 0.29 $ & $( 1.06 )$ \\
${ D }$ {\it vs.} ${ Slope }$ & $  21 $ & $ -0.34 ^{+ 0.28 }_{- 0.23 }$ & $ 0.14 $ & $( 1.49 )$ \\
${ D }$ {\it vs.} ${ a }$ & $  28 $ & $ 0.39 ^{+ 0.18 }_{- 0.22 }$ & $ 0.04 $ & $( 2.06 )$ \\
${ D }$ {\it vs.} ${ e }$ & $  28 $ & $ 0.03 ^{+ 0.23 }_{- 0.23 }$ & $ 0.86 $ & $( 0.17 )$ \\
${ D }$ {\it vs.} ${ i }$ & $  28 $ & $ 0.39 ^{+ 0.16 }_{- 0.19 }$ & $ 0.04 $ & $( 2.06 )$ \\
${ D }$ {\it vs.} ${ q }$ & $  28 $ & $ 0.45 ^{+ 0.15 }_{- 0.19 }$ & $ 0.02 $ & $( 2.40 )$ \\
${ D }$ {\it vs.} ${ Q }$ & $  28 $ & $ 0.34 ^{+ 0.19 }_{- 0.22 }$ & $ 0.08 $ & $( 1.76 )$ \\
{\bf D {\it vs.}  r(h)} &   {\bf 28}  & {\bf 0.56$^{+ 0.14 }_{- 0.17 }$} &  {\bf $<$0.01}  & {\bf ( 3.11 )} \\
${ p_V }$ {\it vs.} ${ Slope }$ & $  21 $ & $ 0.38 ^{+ 0.27 }_{- 0.36 }$ & $ 0.09 $ & $( 1.71 )$ \\
${ p_V }$ {\it vs.} ${ H_V }$ & $  27 $ & $ 0.03 ^{+ 0.27 }_{- 0.28 }$ & $ 0.90 $ & $( 0.13 )$ \\
${ p_V }$ {\it vs.} ${ a }$ & $  28 $ & $ 0.14 ^{+ 0.22 }_{- 0.23 }$ & $ 0.47 $ & $( 0.73 )$ \\
${ p_V }$ {\it vs.} ${ e }$ & $  28 $ & $ 0.32 ^{+ 0.20 }_{- 0.23 }$ & $ 0.10 $ & $( 1.66 )$ \\
${ p_V }$ {\it vs.} ${ i }$ & $  28 $ & $ -0.24 ^{+ 0.23 }_{- 0.21 }$ & $ 0.23 $ & $( 1.21 )$ \\
${ p_V }$ {\it vs.} ${ q }$ & $  28 $ & $ -0.01 ^{+ 0.24 }_{- 0.24 }$ & $ 0.96 $ & $( 0.04 )$ \\
${ p_V }$ {\it vs.} ${ Q }$ & $  28 $ & $ 0.24 ^{+ 0.20 }_{- 0.22 }$ & $ 0.22 $ & $( 1.23 )$ \\
${ p_V }$ {\it vs.} ${ r(h) }$ & $  28 $ & $ 0.07 ^{+ 0.25 }_{- 0.26 }$ & $ 0.74 $ & $( 0.34 )$ \\
\hline
\hline
\noalign{\smallskip}
\multicolumn{5}{c}{Centaurs + SDOs} \\
\hline
\hline
${ D }$ {\it vs.} ${ p_V }$ & $  36 $ & $ -0.13 ^{+ 0.24 }_{- 0.23 }$ & $ 0.45 $ & $( 0.76 )$ \\
${ D }$ {\it vs.} ${ Slope }$ & $  27 $ & $ -0.32 ^{+ 0.20 }_{- 0.17 }$ & $ 0.10 $ & $( 1.65 )$ \\
{\bf D } {\it vs.} {\bf a } & {\bf 36}  & {\bf 0.64 $^{+ 0.11 }_{- 0.15 }$} & {\bf $<$0.01} & {\bf ( 4.22 )} \\
${ D }$ {\it vs.} ${ e }$ & $  36 $ & $ 0.34 ^{+ 0.16 }_{- 0.18 }$ & $ 0.04 $ & $( 2.04 )$ \\
{\bf D } {\it vs.} {\bf i } & {\bf 36}  & {\bf 0.50 $^{+ 0.13 }_{- 0.15 }$} & {\bf $<$0.01}  & {\bf ( 3.13 )} \\
{\bf D } {\it vs.} {\bf q } &  {\bf 36}  & {\bf 0.66 $^{+ 0.10 }_{- 0.14 }$} & {\bf $<$0.01}  & {\bf ( 4.34 )} \\
{\bf D } {\it vs.} {\bf Q } & {\bf 36}  & {\bf 0.61 $^{+ 0.12 }_{- 0.16 }$} & {\bf $<$0.01}  & {\bf ( 3.98 )} \\
{\bf D } {\it vs.} {\bf  r(h) } & {\bf 36}  & {\bf 0.72 $^{+ 0.09 }_{- 0.12 }$} & {\bf $<$0.01}  & {\bf ( 5.01 )} \\
${ p_V }$ {\it vs.} ${ Slope }$ & $  27 $ & $ 0.23 ^{+ 0.26 }_{- 0.29 }$ & $ 0.26 $ & $( 1.13 )$ \\
${ p_V }$ {\it vs.} ${ H_V }$ & $  35 $ & $ -0.01 ^{+ 0.22 }_{- 0.22 }$ & $ 0.95 $ & $( 0.06 )$ \\
${ p_V }$ {\it vs.} ${ a }$ & $  36 $ & $ 0.02 ^{+ 0.20 }_{- 0.20 }$ & $ 0.88 $ & $( 0.14 )$ \\
${ p_V }$ {\it vs.} ${ e }$ & $  36 $ & $ 0.11 ^{+ 0.21 }_{- 0.22 }$ & $ 0.52 $ & $( 0.65 )$ \\
${ p_V }$ {\it vs.} ${ i }$ & $  36 $ & $ -0.10 ^{+ 0.21 }_{- 0.20 }$ & $ 0.57 $ & $( 0.56 )$ \\
${ p_V }$ {\it vs.} ${ q }$ & $  36 $ & $ -0.02 ^{+ 0.21 }_{- 0.21 }$ & $ 0.92 $ & $( 0.10 )$ \\
${ p_V }$ {\it vs.} ${ Q }$ & $  36 $ & $ 0.08 ^{+ 0.19 }_{- 0.19 }$ & $ 0.66 $ & $( 0.44 )$ \\
${ p_V }$ {\it vs.} ${ r(h) }$ & $  36 $ & $ 0.07 ^{+ 0.21 }_{- 0.22 }$ & $ 0.67 $ & $( 0.42 )$ \\
\hline
\hline
\noalign{\smallskip}
\multicolumn{5}{c}{Centaurs smaller than 120 km} \\
\hline
\hline
${ D }$ {\it vs.} ${ p_V }$ & $  23 $ & $ -0.11 ^{+ 0.31 }_{- 0.29 }$ & $ 0.62 $ & $( 0.49 )$ \\
${ D }$ {\it vs.} ${ Slope }$ & $  17 $ & $ -0.01 ^{+ 0.35 }_{- 0.34 }$ & $ 0.97 $ & $( 0.04 )$ \\
${ D }$ {\it vs.} ${ a }$ & $  23 $ & $ 0.48 ^{+ 0.18 }_{- 0.23 }$ & $ 0.02 $ & $( 2.29 )$ \\
${ D }$ {\it vs.} ${ e }$ & $  23 $ & $ 0.20 ^{+ 0.25 }_{- 0.27 }$ & $ 0.37 $ & $( 0.89 )$ \\
${ D }$ {\it vs.} ${ i }$ & $  23 $ & $ 0.39 ^{+ 0.18 }_{- 0.21 }$ & $ 0.06 $ & $( 1.84 )$ \\
${ D }$ {\it vs.} ${ q }$ & $  23 $ & $ 0.47 ^{+ 0.16 }_{- 0.20 }$ & $ 0.02 $ & $( 2.29 )$ \\
${ D }$ {\it vs.} ${ Q }$ & $  23 $ & $ 0.43 ^{+ 0.19 }_{- 0.24 }$ & $ 0.04 $ & $( 2.06 )$ \\
${ D }$ {\it vs.} ${ r(h) }$ & $  23 $ & $ 0.55 ^{+ 0.16 }_{- 0.21 }$ & $ 0.01 $ & $( 2.71 )$ \\
${ p_V }$ {\it vs.} ${ Slope }$ & $  17 $ & $ 0.54 ^{+ 0.20 }_{- 0.30 }$ & $ 0.02 $ & $( 2.22 )$ \\
${ p_V }$ {\it vs.} ${ H_V }$ & $  22 $ & $ -0.15 ^{+ 0.32 }_{- 0.29 }$ & $ 0.52 $ & $( 0.65 )$ \\
${ p_V }$ {\it vs.} ${ a }$ & $  23 $ & $ 0.26 ^{+ 0.25 }_{- 0.29 }$ & $ 0.23 $ & $( 1.20 )$ \\
${ p_V }$ {\it vs.} ${ e }$ & $  23 $ & $ 0.22 ^{+ 0.26 }_{- 0.30 }$ & $ 0.31 $ & $( 1.01 )$ \\
${ p_V }$ {\it vs.} ${ i }$ & $  23 $ & $ -0.13 ^{+ 0.27 }_{- 0.25 }$ & $ 0.56 $ & $( 0.59 )$ \\
${ p_V }$ {\it vs.} ${ q }$ & $  23 $ & $ 0.16 ^{+ 0.28 }_{- 0.31 }$ & $ 0.46 $ & $( 0.73 )$ \\
${ p_V }$ {\it vs.} ${ Q }$ & $  23 $ & $ 0.32 ^{+ 0.23 }_{- 0.27 }$ & $ 0.14 $ & $( 1.47 )$ \\
${ p_V }$ {\it vs.} ${ r(h) }$ & $  23 $ & $ 0.16 ^{+ 0.28 }_{- 0.31 }$ & $ 0.48 $ & $( 0.71 )$ \\
\end{tabular}
\end{center}
\end{table*}


\end{document}